\documentclass[a4paper,11pt]{article}
\pdfoutput=1 

\usepackage{jcappub} 

\usepackage[english]{babel}
\usepackage[utf8x]{inputenc}
\usepackage[T1]{fontenc}

\usepackage{amsmath}
\usepackage{graphicx}
\usepackage{multirow}
\usepackage{units}
\usepackage{tabularx}
\usepackage{amssymb}
\usepackage{booktabs}
\usepackage{array}
\usepackage{physics}
\usepackage{dcolumn}
\usepackage{bm}
\usepackage{xspace}

\usepackage[colorinlistoftodos]{todonotes}

\usepackage{lineno}

\newcommand{\nova}{NOvA\xspace}
\newcommand{\fardet}{Far Detector\xspace}
\newcommand{\neardet}{Near Detector\xspace}

\newcommand{\includegr}[2][scale=0.5]{\includegraphics[#1]{#2}}

\newcounter{raffil} 
\newcommand{\Affiliation}[2]{%
    \affiliation[\refstepcounter{raffil}\label{aff:#1}\ref{aff:#1}]{#2}%
}

\newcounter{raddr} 
\newcommand{\Address}[2]{%
    \affiliation[\refstepcounter{raddr}\label{aff:#1}\ref{aff:#1}]{#2}%
}

\NewDocumentCommand{\Author}{ooom}{%
    \author[\ref{aff:#1}\IfValueT{#2}{,\ref{aff:#2}}\IfValueT{#3}{,\ref{aff:#3}}]{#4}
}

\title{Supernova neutrino detection in NOvA}
\arxivnumber{2005.07155}

\collaborationImg{\includegraphics[width=3cm]{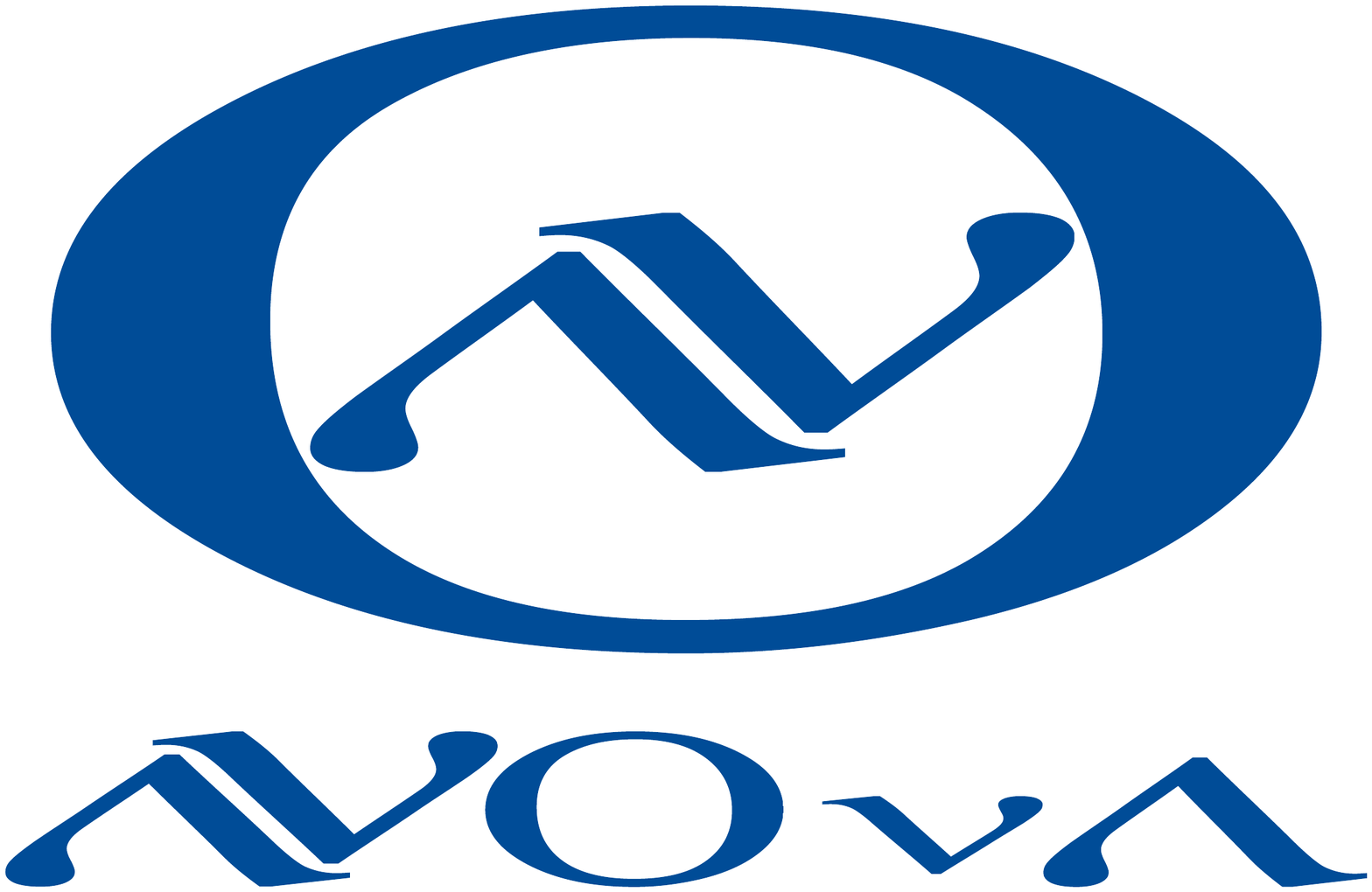}}

\begin{document}

\Author[Atlantico]{M.~A.~Acero}

\Author[FNAL]{P.~Adamson}


\Author[IIT]{G.~Agam}

\Author[FNAL]{L.~Aliaga}

\Author[Sussex]{T.~Alion}

\Author[JINR]{V.~Allakhverdian}




\Author[JINR]{N.~Anfimov}


\Author[JINR]{A.~Antoshkin}


\Author[Magdalena]
{E.~Arrieta-Diaz}

\Author[Sussex]{L.~Asquith}


\Author[Cincinnati]{A.~Aurisano}


\Author[Iowa]{A.~Back}

\Author[Caltech][UCL]{C.~Backhouse}

\Author[Indiana][Sussex][Virginia]{M.~Baird}

\Author[JINR]{N.~Balashov}

\Author[Irvine]{P.~Baldi}

\Author[Hyderabad]{B.~A.~Bambah}

\Author[Tufts]{S.~Bashar}

\Author[Caltech][IIT]{K.~Bays}


\Author[UCL]{S.~Bending}

\Author[FNAL]{R.~Bernstein}


\Author[Panjab]{V.~Bhatnagar}

\Author[Guwahati]{B.~Bhuyan}

\Author[Irvine][Minnesota]{J.~Bian}





\Author[Houston]{J.~Blair}


\Author[Sussex]{A.~C.~Booth}

\Author[CTU]{P.~Bour}



\Author[Indiana]{R.~Bowles}


\Author[MSU]{C.~Bromberg}




\Author[CSU]{N.~Buchanan}

\Author[INR]{A.~Butkevich}

\Author[Minnesota]{V.~Bychkov}

\Author[CSU]{S.~Calvez}




\Author[Texas][Wisconsin]{T.~J.~Carroll}

\Author[Iowa][WandM]{E.~Catano-Mur}



\Author[FNAL]{S.~Childress}

\Author[Delhi]{B.~C.~Choudhary}


\Author[SMU]{T.~E.~Coan}


\Author[WandM]{M.~Colo}


\Author[SDakota]{L.~Corwin}

\Author[UCL]{L.~Cremonesi}



\Author[Mississippi][Indiana]{G.~S.~Davies}




\Author[FNAL]{P.~F.~Derwent}








\Author[FNAL]{P.~Ding}


\Author[ANL]{Z.~Djurcic}

\Author[Tufts]{M.~Dolce}

\Author[CSU]{D.~Doyle}

\Author[Cincinnati]{D.~Due\~nas~Tonguino}

\Author[Virginia]{E.~C.~Dukes}

\Author[Texas]{P.~Dung}

\Author[Carolina]{H.~Duyang}


\Author[Cochin]{S.~Edayath}

\Author[Virginia]{R.~Ehrlich}

\Author[Iowa]{M.~Elkins}

\Author[Harvard]{G.~J.~Feldman}



\Author[IOP]{P.~Filip}

\Author[DallasU]{W.~Flanagan}



\Author[CTU]{J.~Franc}

\Author[SAlabama]{M.~J.~Frank}



\Author[Tufts]{H.~R.~Gallagher}

\Author[MSU]{R.~Gandrajula}

\Author[Pitt]{F.~Gao}

\Author[UCL]{S.~Germani}




\Author[IHyderabad]{A.~Giri}


\Author[UFG]{R.~A.~Gomes}


\Author[ANL]{M.~C.~Goodman}

\Author[Lebedev]{V.~Grichine}

\Author[Indiana]{M.~Groh}


\Author[Virginia]{R.~Group}




\Author[Carolina]{B.~Guo}

\Author[Duluth]{A.~Habig}

\Author[ICS]{F.~Hakl}

\Author[Virginia]{A.~Hall}


\Author[Sussex]{J.~Hartnell}

\Author[FNAL]{R.~Hatcher}

\Author[Tennessee]{A.~Hatzikoutelis}

\Author[Minnesota]{K.~Heller}

\Author[Cincinnati]{J.~Hewes}

\Author[FNAL]{A.~Himmel}

\Author[UCL]{A.~Holin}

\Author[Indiana]{B.~Howard}

\Author[Texas]{J.~Huang}




\Author[FNAL]{J.~Hylen}


\Author[CTU]{F.~Jediny}





\Author[CSU]{C.~Johnson}


\Author[CSU]{M.~Judah}


\Author[JINR]{I.~Kakorin}

\Author[Panjab]{D.~Kalra}


\Author[IIT]{D.~M.~Kaplan}



\Author[Cochin]{R.~Keloth}


\Author[JINR]{O.~Klimov}

\Author[Houston]{L.~W.~Koerner}


\Author[JINR]{L.~Kolupaeva}

\Author[Lebedev]{S.~Kotelnikov}





\Author[CTU]{M.~Kubu}

\Author[JINR]{Ch.~Kullenberg}

\Author[Panjab]{A.~Kumar}


\Author[Carolina]{C.~D.~Kuruppu}

\Author[CTU]{V.~Kus}




\Author[Indiana]{T.~Lackey}


\Author[Texas]{K.~Lang}






\Author[Irvine]{L.~Li}

\Author[CSU]{S.~Lin}

\Author[Wisconsin]{A.~Lister}


\Author[IOP]{M.~Lokajicek}




\Author[INR]{S.~Luchuk}




\Author[ANL]{S.~Magill}

\Author[Tufts]{W.~A.~Mann}

\Author[Minnesota]{M.~L.~Marshak}



\Author[Iowa]{M.~Martinez-Casales}




\Author[INR]{V.~Matveev}


\Author[Sussex]{B.~Mayes}



\Author[Sussex]{D.~P.~M\'endez}


\Author[Indiana]{M.~D.~Messier}

\Author[WSU]{H.~Meyer}

\Author[FNAL]{T.~Miao}



\Author[Minnesota]{W.~H.~Miller}

\Author[Carolina]{S.~R.~Mishra}

\Author[Minnesota]{A.~Mislivec}

\Author[Hyderabad]{R.~Mohanta}

\Author[Duluth]{A.~Moren}

\Author[JINR]{A.~Morozova}

\Author[Caltech]{L.~Mualem}

\Author[WSU]{M.~Muether}

\Author[Indiana]{S.~Mufson}

\Author[UCL]{K.~Mulder}

\Author[Indiana]{R.~Murphy}

\Author[Indiana]{J.~Musser}

\Author[Pitt]{D.~Naples}

\Author[Irvine]{N.~Nayak}


\Author[WandM]{J.~K.~Nelson}

\Author[UCL]{R.~Nichol}

\Author[IIT]{G.~Nikseresht}

\Author[Indiana][FNAL]{E.~Niner}

\Author[FNAL]{A.~Norman}

\Author[FNAL]{A.~Norrick}

\Author[Charles]{T.~Nosek}



\Author[JINR]{A.~Olshevskiy}


\Author[Tufts]{T.~Olson}

\Author[FNAL]{J.~Paley}



\Author[Caltech]{R.~B.~Patterson}

\Author[Minnesota]{G.~Pawloski}




\Author[JINR]{O.~Petrova}


\Author[Carolina]{R.~Petti}





\Author[FNAL]{R.~K.~Plunkett}




\Author[Indiana][Texas]{F.~Psihas}





\Author[ANL]{A.~Rafique}

\Author[Caltech]{V.~Raj}


\Author[FNAL]{B.~Ramson}


\Author[FNAL][Wisconsin]{B.~Rebel}





\Author[CSU]{P.~Rojas}




\Author[Lebedev]{V.~Ryabov}





\Author[JINR]{O.~Samoylov}

\Author[Iowa][ANL]{M.~C.~Sanchez}

\Author[Iowa]{S.~S\'{a}nchez~Falero}





\Author[Irvine]{I.~S.~Seong}


\Author[FNAL]{P.~Shanahan}



\Author[JINR][Corresponding]{A.~Sheshukov}



\Author[Delhi]{P.~Singh}

\Author[BHU]{V.~Singh}



\Author[Indiana]{E.~Smith}

\Author[CTU]{J.~Smolik}

\Author[IIT]{P.~Snopok}

\Author[WSU]{N.~Solomey}



\Author[Cincinnati]{A.~Sousa}

\Author[Charles]{K.~Soustruznik}


\Author[Minnesota]{M.~Strait}

\Author[ANL][FNAL]{L.~Suter}

\Author[Virginia]{A.~Sutton}

\Author[UCL]{C.~Sweeney}

\Author[ANL]{R.~L.~Talaga}


\Author[Texas]{B.~Tapia~Oregui}


\Author[Charles]{P.~Tas}


\Author[Cochin]{R.~B.~Thayyullathil}

\Author[UCL][Wisconsin]{J.~Thomas}



\Author[Iowa]{E.~Tiras}




\Author[Minnesota]{D.~Torbunov}


\Author[Panjab]{J.~Tripathi}

\Author[FNAL]{A.~Tsaris}

\Author[IIT]{Y.~Torun}


\Author[Indiana]{J.~Urheim}

\Author[WandM]{P.~Vahle}

\Author[Caltech]{Z.~Vallari}

\Author[Indiana]{J.~Vasel}



\Author[CTU]{P.~Vokac}


\Author[CTU]{T.~Vrba}


\Author[Cincinnati]{M.~Wallbank}



\Author[Iowa]{T.~K.~Warburton}



\Author[Iowa]{M.~Wetstein}


\Author[Syracuse][Indiana]{D.~Whittington}

\Author[FNAL]{D.~A.~Wickremasinghe}





\Author[Stanford]{S.~G.~Wojcicki}

\Author[Tufts]{J.~Wolcott}





\Author[Syracuse]{A.~Yallappa~Dombara}


\Author[FNAL]{K.~Yonehara}

\Author[ANL][IIT]{S.~Yu}

\Author[IIT]{Y.~Yu}

\Author[INR]{S.~Zadorozhnyy}

\Author[IOP]{J.~Zalesak}


\Author[Sussex]{Y.~Zhang}



\Author[FNAL]{R.~Zwaska}

\collaboration{The NOvA Collaboration}

\Affiliation{ANL}{Argonne National Laboratory, Argonne, Illinois 60439, USA}
\Affiliation{Atlantico}{Universidad del Atlantico, Carrera 30 No. 8-49, Puerto Colombia, Atlantico, Colombia}
\Affiliation{BHU}{Department of Physics, Institute of Science, Banaras Hindu University, Varanasi, 221 005, India}
\Affiliation{Caltech}{California Institute of Technology, Pasadena, California 91125, USA}
\Affiliation{Charles} {Charles University, Faculty of Mathematics and Physics, Institute of Particle and Nuclear Physics, Prague, Czech Republic}
\Affiliation{Cincinnati}{Department of Physics, University of Cincinnati, Cincinnati, Ohio 45221, USA}
\Affiliation{Cochin}{Department of Physics, Cochin University of Science and Technology, Kochi 682 022, India}
\Affiliation{CSU}{Department of Physics, Colorado State University, Fort Collins, CO 80523-1875, USA}
\Affiliation{CTU}{Czech Technical University in Prague, Brehova 7, 115 19 Prague 1, Czech Republic}
\Affiliation{DallasU}{University of Dallas, 1845 E Northgate Drive, Irving, Texas 75062 USA}
\Affiliation{Delhi}{Department of Physics and Astrophysics, University of Delhi, Delhi 110007, India}
\Affiliation{FNAL}{Fermi National Accelerator Laboratory, Batavia, Illinois 60510, USA}
\Affiliation{UFG}{Instituto de F\'{i}sica, Universidade Federal de Goi\'{a}s, Goi\^{a}nia, Goi\'{a}s, 74690-900, Brazil}
\Affiliation{Guwahati}{Department of Physics, IIT Guwahati, Guwahati, 781 039, India}
\Affiliation{Harvard}{Department of Physics, Harvard University, Cambridge, Massachusetts 02138, USA}
\Affiliation{Houston}{Department of Physics, University of Houston, Houston, Texas 77204, USA}
\Affiliation{IHyderabad}{Department of Physics, IIT Hyderabad, Hyderabad, 502 205, India}
\Affiliation{Hyderabad}{School of Physics, University of Hyderabad, Hyderabad, 500 046, India}
\Affiliation{IIT}{Illinois Institute of Technology, Chicago IL 60616, USA}
\Affiliation{Indiana}{Indiana University, Bloomington, Indiana 47405, USA}
\Affiliation{ICS}{Institute of Computer Science, The Czech Academy of Sciences, 182 07 Prague, Czech Republic}
\Affiliation{INR}{Institute for Nuclear Research of Russia, Academy of 
Sciences 7a, 60th October Anniversary prospect, Moscow 117312, Russia}
\Affiliation{IOP}{Institute of Physics, The Czech Academy of Sciences, 182 21 Prague, Czech Republic}
\Affiliation{Iowa}{Department of Physics and Astronomy, Iowa State University, Ames, Iowa 50011, USA}
\Affiliation{Irvine}{Department of Physics and Astronomy, University of California at Irvine, Irvine, California 92697, USA}
\Affiliation{JINR}{Joint Institute for Nuclear Research,  Dubna, Moscow region 141980, Russia}
\Affiliation{Lebedev}{Nuclear Physics and Astrophysics Division, Lebedev Physical Institute, Leninsky Prospect 53, 119991 Moscow, Russia}
\Affiliation{Magdalena}{Universidad del Magdalena, Carrera 32 No 22 – 08 Santa Marta, Colombia}
\Affiliation{MSU}{Department of Physics and Astronomy, Michigan State University, East Lansing, Michigan 48824, USA}
\Affiliation{Duluth}{Department of Physics and Astronomy, University of Minnesota Duluth, Duluth, Minnesota 55812, USA}
\Affiliation{Minnesota}{School of Physics and Astronomy, University of Minnesota Twin Cities, Minneapolis, Minnesota 55455, USA}
\Affiliation{Mississippi}{University of Mississippi, University, Mississippi 38677, USA}
\Affiliation{Panjab}{Department of Physics, Panjab University, Chandigarh, 160 014, India}
\Affiliation{Pitt}{Department of Physics, University of Pittsburgh, Pittsburgh, Pennsylvania 15260, USA}
\Affiliation{SAlabama}{Department of Physics, University of South Alabama, Mobile, Alabama 36688, USA} 
\Affiliation{Carolina}{Department of Physics and Astronomy, University of South Carolina, Columbia, South Carolina 29208, USA}
\Affiliation{SDakota}{South Dakota School of Mines and Technology, Rapid City, South Dakota 57701, USA}
\Affiliation{SMU}{Department of Physics, Southern Methodist University, Dallas, Texas 75275, USA}
\Affiliation{Stanford}{Department of Physics, Stanford University, Stanford, California 94305, USA}
\Affiliation{Sussex}{Department of Physics and Astronomy, University of Sussex, Falmer, Brighton BN1 9QH, United Kingdom}
\Affiliation{Syracuse}{Department of Physics, Syracuse University, Syracuse NY 13210, USA}
\Affiliation{Tennessee}{Department of Physics and Astronomy, University of Tennessee, Knoxville, Tennessee 37996, USA}
\Affiliation{Texas}{Department of Physics, University of Texas at Austin, Austin, Texas 78712, USA}
\Affiliation{Tufts}{Department of Physics and Astronomy, Tufts University, Medford, Massachusetts 02155, USA}
\Affiliation{UCL}{Physics and Astronomy Dept., University College London, Gower Street, London WC1E 6BT, United Kingdom}
\Affiliation{Virginia}{Department of Physics, University of Virginia, Charlottesville, Virginia 22904, USA}
\Affiliation{WSU}{Department of Mathematics, Statistics, and Physics, Wichita State University, Wichita, Kansas 67206, USA}
\Affiliation{WandM}{Department of Physics, William \& Mary, Williamsburg, Virginia 23187, USA}
\Affiliation{Wisconsin}{Department of Physics, University of Wisconsin-Madison, Madison, Wisconsin 53706, USA}
\Address{Corresponding}{Corresponding Author: andrey.sheshukov@jinr.ru}
\newcommand{\deceased}{Deceased.}


 \begin{abstract} {
  The NOvA long-baseline neutrino experiment uses a pair of large,
  segmented, liquid-scintillator calorimeters to study neutrino
  oscillations, using GeV-scale neutrinos from the Fermilab NuMI beam. These detectors are also sensitive to the flux of
  neutrinos which are emitted during a core-collapse supernova through inverse beta decay interactions on carbon at energies of $\mathcal{O}(10~\text{MeV})$.  This signature provides a means to study the dominant mode
  of energy release for a core-collapse supernova occurring in our
  galaxy.  We describe the data-driven software trigger system
  developed and employed by the NOvA experiment to identify and record
  neutrino data from nearby galactic supernovae. This technique has been used by \nova{} 
  to self-trigger on potential core-collapse supernovae in our galaxy, with an estimated sensitivity 
reaching out to \unit[10]{kpc} distance while achieving a detection efficiency
of 23\% to 49\% for supernovae from progenitor
stars with masses of \unit[9.6]{M$_\odot$} to \unit[27]{M$_\odot$}, respectively.
}
\end{abstract}

\maketitle
\flushbottom

\section{Introduction}
The modeling and understanding of the stellar dynamics involved in
core-collapse supernovae (SN) events require knowledge of the complex
interplay between different physics processes which transition rapidly
during the initial collapse of the star and the explosive expansion
phases of the event.  The neutrino burst that drives this explosive
phase has been observed in the SN1987a event~\cite{Bionta:1987qt,Hirata:1987hu,Alekseev:1987ej,Aglietta:1987it}, but detailed
measurement of the neutrino energy spectrum and time evolution of the
neutrino flux are central to our understanding of these processes and will
allow for more detailed models of the stellar dynamics to be
evaluated.  

The NOvA data acquisition
system has been designed to detect and trigger on the time evolution of the rates of low energy
interactions within the detectors.  These interaction rates are 
compared, as part of the triggering, to the corresponding low energy interaction rates that are  expected from models 
of a core-collapse supernova in our galaxy.  This triggering process
allows the \nova{} experiment to capture both the energy spectrum
and time evolution of the neutrino flux, 
contributing to the world's dataset on the dynamics of supernovae.

These trigger-level identification decisions allow data to be saved
from each detector's data acquisition system's memory buffers to permanent storage for later
offline analysis.  The trigger decisions are also designed to be sent
as alerts to the Supernova Early Warning System (SNEWS)~\cite{snews}
coincidence network where they can be combined with alerts and
observations from other neutrino experiments worldwide.  Because the
expected rate of nearby supernovae is estimated at only a few per
century, the trigger we present has been specifically tuned to
preserve signal triggering efficiency at the cost of a higher rate of
false positive trigger events.  The trigger presented here, tuned to detector performance data, is projected to have an
average false positive rate, for the \nova{} \fardet operating on the
surface with minimal shielding and overburden, based on Poisson
fluctuations of noise and cosmic ray induced activity, of one event
per 7 days.

This trigger chain is discussed in the section \ref{sec:detectors} that follows.
Section~\ref{sec:sim} describes the newly developed supernova flux
generation scheme required to propagate the model dependent energy and
time structure of neutrino interactions into the full detector
response simulation.  Section~\ref{sec:selection} then describes how
data collected by the detectors are filtered and reconstructed in real
time to suppress known backgrounds and identify potential low energy
neutrino interactions.  The methodology and implementation of the
supernova trigger system within the NOvA detectors' readout and data
acquisition system are presented in Sec.~\ref{sec:ddt}. Details of the
system's performance and resulting detection sensitivities are
reported and summarized in Sec.~\ref{sec:sensitiv}.  The results and
statistics of the trigger performance during the commissioning
period are presented in Sec.~\ref{sec:commissioning}.

\subsection{NOvA Detectors}
\label{sec:detectors}
NOvA is a long-baseline neutrino oscillation
experiment~\cite{nova-tdr} using a pair of functionally identical
liquid scintillator calorimeters to study electron neutrino appearance
in the primarily muon neutrino NuMI beam~\cite{numi_nim} with a central
energy of 
\unit[2]{GeV}.  The \unit[14]{kt} \fardet (FD)
is located on the surface at the Ash River site in northern Minnesota,
USA and is shielded
by a concrete roof topped by an additional \unit[15]{cm} of barite stone. The \unit[810]{km} baseline for the neutrino oscillation measurements is determined by the separation between this site and the beam source. The approximately \unit[300]{t} 
Near Detector
(ND) is situated \unit[100]{m} underground at Fermilab, \unit[1]{km} from the
beam source.  Both detectors
are comprised of planes of extruded polyvinyl chloride (PVC) composed
of a custom formulation with titanium dioxide to increase
reflectivity~\cite{extrusion_nim}.  Each extrusion panel is segmented into
\unit[3.9]{cm} by \unit[6.6]{cm} detection cells. The lengths of these
detection cells are \unit[15.5]{m} for the \fardet, while the shorter
\neardet cells are \unit[3.9]{m} long. The detector planes are
constructed from these extruded panels resulting in 384 cell wide
planes (FD) or 96 cell wide planes (ND).  Each cell is filled with a
mineral oil based liquid scintillator~\cite{scint_nim}, resulting in a 63\% active mass fraction.
The planes are alternated between a vertical (X-measuring) and
horizontal (Y-measuring) orientation, with the Z-axis taken to be
orthogonal to these planes and aligned with the beam direction.  This
orthogonal arrangement allows for simultaneous reconstruction of
particle trajectories in the projected X/Z and Y/Z views as well as
for three dimensional reconstruction through associative matching of
hit information between readout views.

Scintillation light is collected and transported out of each cell via
a loop of wavelength shifting fiber.  This light is detected by
an avalanche photodiode (APD) and the signal digitized by high speed
readout electronics, providing both charge and time 
information for each of the 344,064 cells of the \fardet at an
effective continuous digitization rate of
\unit[2]{MHz}~\cite{Norman:CHEP2015-6}.  The resulting hit data is
collated by the acquisition system's front-end readout
electronics, and then transmitted to and buffered in the distributed
memory of the experiment’s trigger and buffer farm.  This readout and
buffer scheme provides circular style memory buffers which can be
examined in real time to make triggering decisions or used by the
final stages of event building to record portions of the data to
permanent storage~\cite{Norman:CHEP2015-6}.  The same is done for the
\neardet, but the number of channels (20,192) is smaller and the  effective sampling rate is larger (\unit[8]{MHz}).

The amplitude of the light signal is digitized in ADC counts, providing a measure of the energy deposited in the cell. For example, a muon crossing the center of a cell deposits about \unit[10]{MeV} producing around \unit[240]{ADC} counts in the \fardet, or \unit[530]{ADC} counts in the smaller \neardet. 

The raw data rate into the circular buffer systems averages
$\sim$\unit[1.2]{GB/s} at the \fardet under normal operating
conditions.  This rate is nearly constant and dominated by the
more than \unit[100]{kHz} rate of cosmic ray muons that penetrate the \fardet
 overburden.  This data stream is broken into \unit[5]{ms} time
windows, referred to as ``milliblocks'', which are stored in RAM
buffers of a farm of commodity servers that are part of the Data Acquisition System (DAQ).  The buffer farm has enough memory to store 
\unit[1350]{s} of data.  This results in the ability of the system to
look backwards in time up to the depth of the buffer for
identifying supernova signatures. The same methodology is used at the
\neardet, although its smaller size and larger
overburden reduces the data rate to $\sim$\unit[13]{MB/s} and
increases the buffer depth to \unit[1900]{s} despite having fewer buffer nodes in the acquisition pool.

When the DAQ and trigger systems identifies a
signature of interest, it designates a time window of data to record.
The window is variable and can range from a minimum duration of
\unit[50]{$\mu$s} to the maximum duration corresponding to the size of
the buffer.  This data is retrieved from the buffers, collated,
assembled into a custom data format, and written to disk for later
analysis.  Since the servers housing the buffer RAM also have
significant CPU resources, the buffered data are examined
asynchronously for patterns and topologies of interest.  When a
pattern or topology is identified, a trigger decision is issued which
causes the corresponding data to be saved from the buffers to the NOvA
data streams.  This process of data-driven triggering is used to
identify and record the elevated rates of small hit clusters that are
indicative of a supernova neutrino burst, and has been operational
since November~1, 2017.

\subsection{Supernova neutrino signal}

Core-collapse supernovae produce around $10^{58}$ neutrinos during the
first seconds after the explosion. These neutrinos carry away about
$99\%$ of the gravitational binding energy released by the collapse
and play a crucial role in the explosion mechanism of the
star~\cite{sn-review}.  In particular, during the first \unit[10]{ms}
of the supernova, when the stellar core collapses to a neutron star,
$1\%$ of the total neutrino flux is produced as electron neutrinos due
to the neutronization process.  The enormous neutrino density
interacts with the collapsing matter to power the shockwave
that triggers the supernova explosion.  This process is
subsequently accompanied by an emission of neutrinos and anti-neutrinos
of all flavors which are created through the pair-production channel.
Because of their low interaction cross section, neutrinos can escape
easily compared to more strongly interacting particles, which instead
recycle their energy back into the hot dense medium of the
proto-neutron star.  These neutrino emissions serve as the dominant
cooling mechanism for the proto-neutron star.

Modern models of core-collapse supernova explosions often feature
complex 3D simulations of the processes during the core collapse.
Results of these simulations predict various neutrino spectra and
energy-luminance curves vs. time for different models that probe both
astrophysics and neutrino parameters
\cite{livermore,garching,sndb}. Aside from these subtleties, the basic
features of the neutrino ``light curve'' are common to all models.  The light curve exhibits 
an initial, short neutronization burst, then there is about a 1 second of neutrino emissions as the proto-neutron star accretes infalling matter and the shock front develops.  
This is followed by the emissions during
exponential cooling, which occur over the next tens of seconds.  The
neutrinos, prior to any oscillation or collective effects, are
produced in roughly equal populations with respect to lepton
flavor.  These populations have energy spectra in the tens of MeV, with
slightly hotter spectra for the muon and tau neutrino species, cooler for electron anti-neutrinos,  and the
coolest for the electron neutrinos.

\begin{figure*}[!t]
\centerline{
\includegr{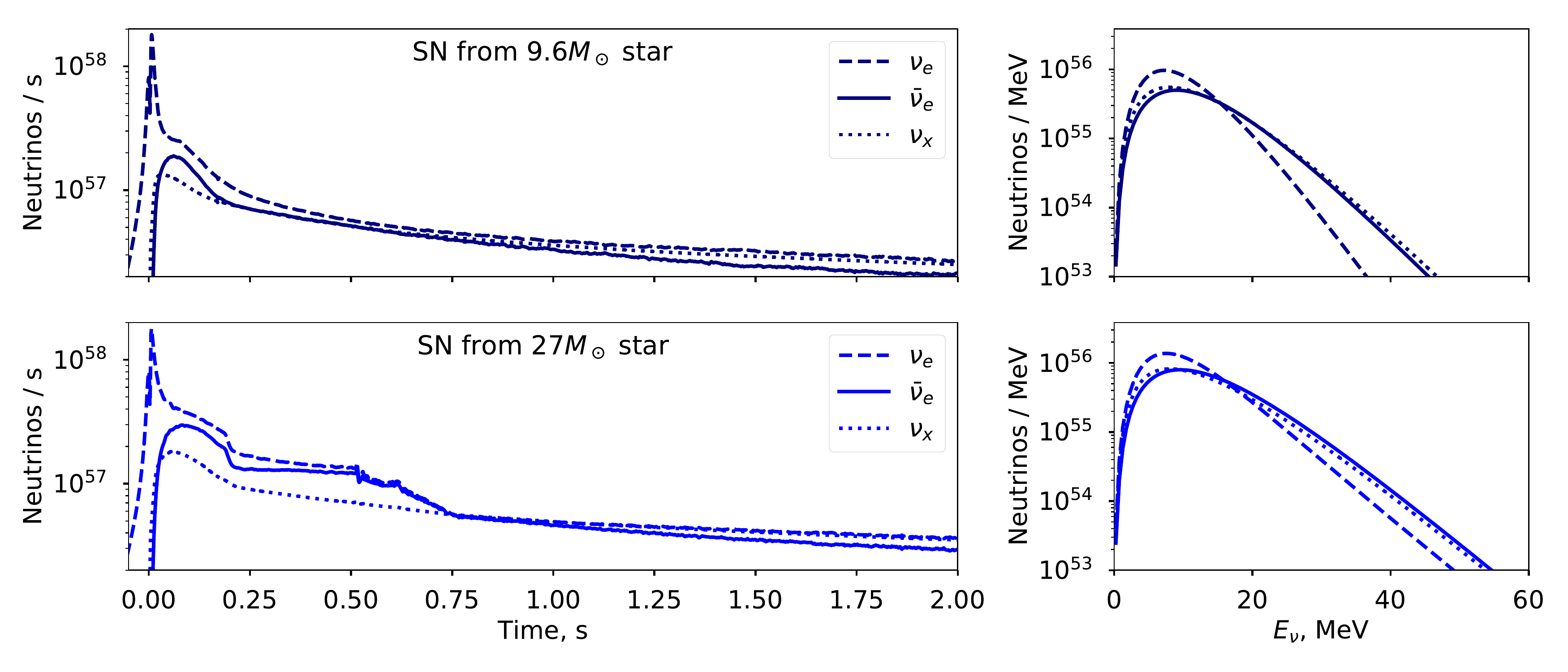}
}
\caption{\label{fig:nu_lumi} Expected neutrino production vs. time
  (left) and energy (right) from collapsing stars with a mass of
  \unit[9.6]{M$_\odot$} (top) and \unit[27]{M$_\odot$} (bottom), from
  the simulation by the Garching group \cite{garching}. This
  simulation does not include flavor changing effects such as neutrino
  oscillations and collective effects.  }
\end{figure*}

In this work we use the neutrino flux from the simulations by the
Garching group~\cite{garching}.  This is done for two progenitor star
masses of $9.6$ and $27$ solar masses (see Fig.~\ref{fig:nu_lumi}), which the Garching group have chosen as representative of typical low- and high-mass supernovae.  
We do not consider the effects of neutrino oscillations or
self-interactions in this work, for purposes of more straightforward comparisons to other detectors (as per Ref.~\cite{sn-review}), although Sec.~\ref{ssec:results} describes their rough effect on the triggering. These effects, potentially leading to
spectral distortions, ``swapping'', or ``splits'' (see Sec. 4 in
\cite{garching}), could provide the opportunity to study the neutrino
properties from an observed supernova neutrino signal.  NOvA's sensitivities to such interesting physics will be the subject of a future paper.

\subsection {Detection channels}

\begin{figure}[!t]
\centering
\includegr{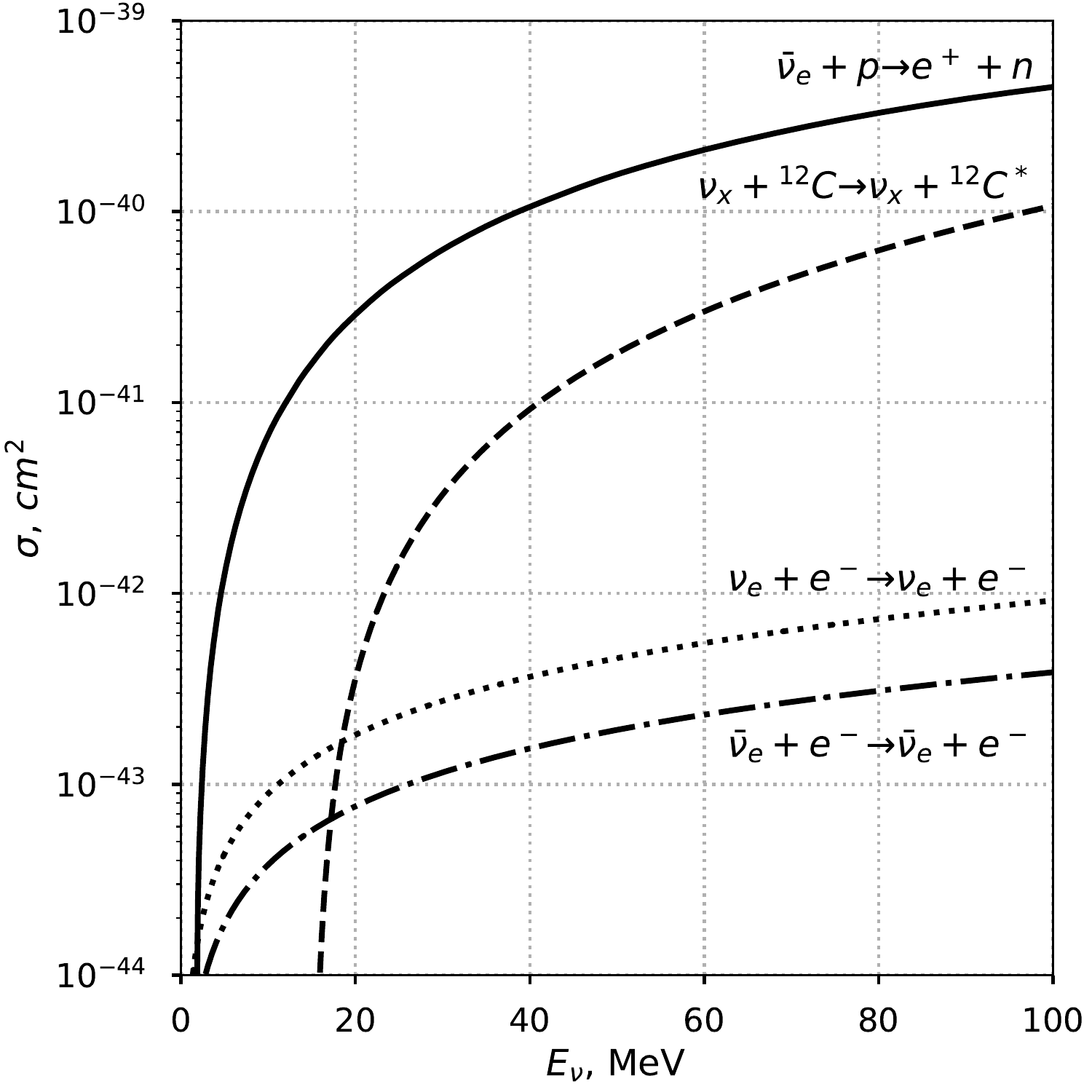}
\caption{\label{fig:xsec} Cross section vs. energy for the neutrino interaction channels relevant for the NOvA detectors, from \cite{StrumiaVissani,NuElastic,NConC12}.}
\end{figure}

The observation of the neutrino signal from a core-collapse supernova
is highly dependent on the detector's technology and composition.  In the case of a hydrocarbon
based scintillator detector like those used in the NOvA experiment,
the prominent channels for observation of neutrinos in the tens of MeV
range are inverse beta decay (IBD), elastic scattering on
electrons, and neutral current interactions on carbon (see
Fig.~\ref{fig:xsec}).  These signatures, as observable in NOvA, are
described in the following subsections and summarized in Tab.~\ref{tab:sigrates}, while other interaction channels produce few observable signals in the NOvA detectors.

\subsubsection{Inverse beta decay (IBD)}
$$
\bar{\nu}_e+p\to e^++n
$$ 
When an electron anti-neutrino interacts with a proton in the low-Z
hydrocarbons which make up both the active and inactive regions of the
NOvA detectors, it produces a positron and a low energy neutron.  Both
of these particles can propagate significant distances within the
active regions of the detector and register as activity.  To
compute the IBD reaction rates in the NOvA detector, we use the IBD
cross-section calculation by Strumia and Vissani
\mbox{\cite[Table 1]{StrumiaVissani}}, as it applies specifically to supernova
neutrino energies.  This interaction channel is by far the dominant
channel for supernova bursts that can be detected in the near and far
NOvA detectors.  In addition, there is a simple kinematic correlation
between the energy of the incident anti-neutrino and the energy of the
resulting positron, which is observed in the final state of the
interaction.  In this manner, the IBD channel allows for a computation
of the neutrino energy spectrum  from the positron kinetic energy $E_{e^+}$ via
\begin{equation}
  E_{\bar{\nu}_e} \simeq E_{e^+} + \text{1.29~MeV}.
\end{equation}

\subsubsection{Elastic scattering of neutrinos on electrons}
$$
{\nu}_x(\bar{\nu}_x)+e^-\to {\nu}_x(\bar{\nu}_x)+e^-
$$
The process of elastic scattering of neutrinos on electrons is common
to all (anti)neutrino flavors \cite{NuElastic}.  In the final
state the scattered electron is observed.  The observed amplitude for
the process must be summed over the active neutrino and anti-neutrino
species.  However, the $\nu_e$-electron scattering has the
largest interaction cross section and is the dominant interaction mode
for this process.  In the NOvA detector, this results in a larger
contribution to the observed activity coming from the $\nu_e$ channel
as compared with the other five subdominant interaction channels.  Unlike the IBD channel, this process preserves significant
information on the directionality of the incident neutrino, which is
carried through into the final state kinematics of the observed
electron.  However, in the case of the NOvA detector geometry, the
resulting electrons are low enough in energy that their mean path
length through the active detector mass when combined with the coarse
($\unit[3.9]{cm}\times\unit[6.6]{cm}$) granularity of the NOvA readout cells is insufficient to
reconstruct the incident neutrino angle with the
precision needed to provide pointing information for the parent supernova.
As a result, this channel is included in the supernova signal rate
calculations, but no attempt is made to reconstruct neutrino
directions from it.

\subsubsection{Neutral current scattering on carbon}
$$
{\nu}_x(\bar{\nu}_x)+{}^{12}C\to {\nu}_x(\bar{\nu}_x)+{}^{12}C^*(15.1\text{ MeV})
$$
Neutral current neutrino interactions probe the full flavor
content of the neutrino sector, permitting the $\nu_\mu$ and
$\nu_\tau$ components of the supernova to be detected, unlike the
corresponding charged current interactions which have interaction thresholds
significantly above the modeled supernova neutrino energy spectra.
In particular, neutral current interactions on carbon nuclei result in a nuclear excitation.  The subsequent \unit[15.1]{MeV} de-excitation photon~\cite{NConC12} is observable as near-threshold activity in the NOvA detectors, with higher efficiency in the \neardet with its shorter cells and consequent smaller light attenuation.  
While this process is subdominant compared to the elastic scattering
and IBD channels, it does contribute to the overall rate and provides
a mechanism to probe the interactions independent of lepton flavor.

\begin{table}[!t]
  \centering
  \caption{\label{tab:sigrates}Estimated average number of neutrino interactions in the NOvA detectors for dominant interaction and detection channels for Garching supernova neutrino flux simulations from \unit[9.6]{$M_\odot$} and \unit[27]{$M_\odot$} progenitor stars at a distance of \unit[10]{kpc}.}
\begin{tabularx}{\linewidth}{Xrr@{\hskip 5pt}rr}
\toprule
\multirow{2}{*}{Interaction channel} & \multicolumn{2}{c}{\fardet}\hskip 5pt & \multicolumn{2}{c}{\neardet}\\
&  $\unit[9.6]{M_\odot}$ & $\unit[27]{M_\odot}$ &     $\unit[9.6]{M_\odot}$ &  $\unit[27]{M_\odot}$ \\
\midrule
Inverse beta decay                     & 1593 & 3439 & 24 & 51 \\
Elastic scattering on $e^{-}$         &  143 &  259 &  3 &  5 \\
Neutral current on ${}^{12}\mathrm{C}$ &   67 &  166 &  1 &  3 \\
\midrule
Total & 1803 & 3864 &    28 & 59 \\
\bottomrule
\end{tabularx}
\end{table}

\section{Simulation chain}
\label{sec:sim}

Monte Carlo simulations of supernova neutrino bursts are
performed in order to understand how the resulting signal channels
will register within the NOvA detectors.  These simulations are
produced using different models for the predicted supernova fluence,
and the resulting predictions are then scaled by the distance to the
supernova to understand the signal rates.

These fluences are convolved with the neutrino
interaction cross sections on the NOvA detector materials to predict
the expected interaction rates within the detector.  This convolution
is performed on an interaction-by-interaction basis so that detailed
examination of the final state particles of each interaction can be
propagated through the detector simulation.  The detector
simulation incorporates both detailed models both for the particle
propagation through the media as well as for light yields and
propagation of scintillation and Cherenkov light through the detector.
The simulation also includes custom models of the NOvA electronics
and digitization chain, which utilize measured detector
responses and detector efficiencies.  The combination of steps in this end-to-end simulation chain thus allows the
computation of the detection efficiencies for the supernova
signal channels, as well as the overall efficiencies for
supernova trigger algorithms which operate on the ensemble of
signal and background activity.

\subsection{GenieSNova}

\begin{figure*}[!t]
    \centerline{
    \includegr{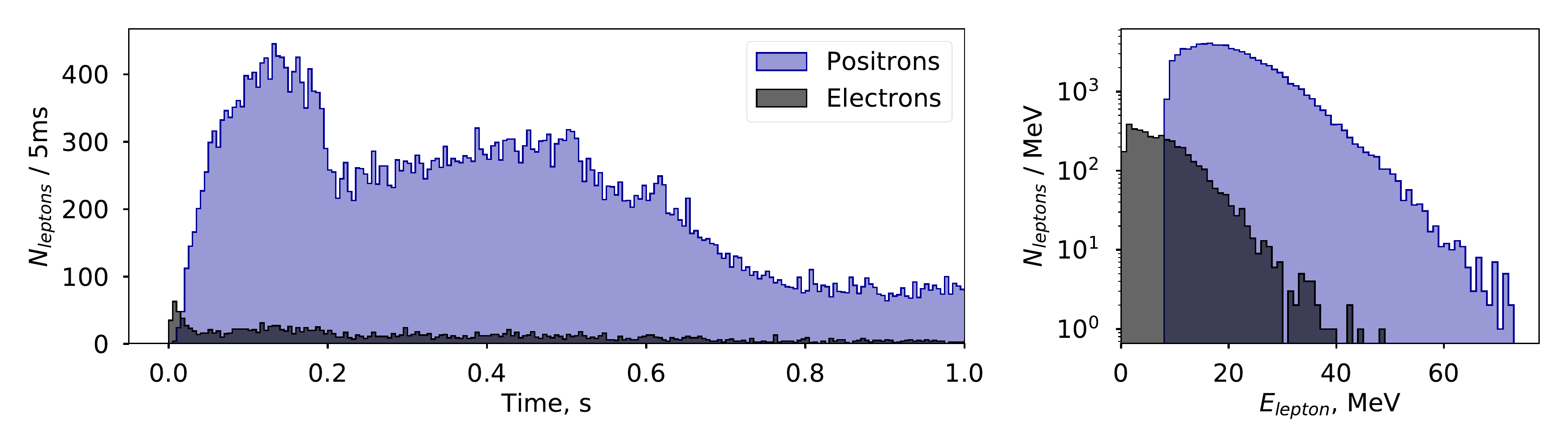}
    }
    \caption{Results of GenieSNova simulation of SN neutrinos from a
      \unit[27]{$M_\odot$} progenitor star at \unit[2]{kpc} distance,
      interacting in the NOvA \fardet: time (left) and energy (right)
      distributions of secondary charged leptons. The electrons come
      from the neutrino-electron elastic scattering, and positrons are
      produced in the IBD process. The minimum neutrino energy for
      this simulation was set to \unit[10]{MeV} to reduce the
      calculation time, which leads to a cutoff in the positron
      spectrum, seen on the right plot. This does not affect the NOvA
      result, as positrons below \unit[10]{MeV} are not reconstructed
      by NOvA (see Fig.~\ref{fig:posit_efficiency}).  }
    \label{fig:gsnova}
\end{figure*}

For the NOvA supernova trigger, a dedicated software package named
``GenieSNova'' was developed to simulate interactions of supernova
neutrinos inside the NOvA detectors.  This new package generates
neutrinos according to a supernova neutrino flux distribution including both the time structure and energy distributions.  It then then simulates neutrino interactions on the NOvA
materials according to the detectors' geometry.  The GenieSNova package
interfaces models of supernova neutrino fluence from the Garching
group~\cite{garching} and the Supernova Neutrino Database~\cite{sndb}
with the GENIE neutrino interactions generator \cite{genie} to
accomplish this simulation.

The daughter particles from the interactions are then propagated
through the detector with the Geant4~\cite{Geant} based NOvA detector
simulation~\cite{nova-detsim} which is used throughout the NOvA
experiment for both beam neutrino analysis and non-beam studies.  This
simulation chain allows for easy integration of the supernova analysis
into the standard NOvA simulation pipelines and permits the use of
standard NOvA reconstruction and analysis tools for clustering,
tracking, and event reconstruction.

The development of GenieSNova was required to properly simulate the
macroscopic time structure of the supernova neutrino signals under different model
assumptions.  The standard GENIE generator has been developed
primarily for accelerator beam induced interactions and for
atmospheric neutrino interactions.  Both of these neutrino sources are not considered to have extended time correlations.  In contrast, the supernova neutrino burst models required correlation of the neutrino generation across time scales on the order of tens of seconds. To develop a trigger which relies on the time structure of the burst, the ability to produce both time and energy dependent neutrino interactions had to be introduced to GENIE.  In addition to the extended time structures, the GenieSNova
simulation modified the NOvA interfaces to the underlying GENIE generator to allow for neutrino energies down to \unit[1]{MeV}  to be correctly propagated through the simulation infrastructure.  This is in contrast to the nominal beam neutrino simulation that is used for the NOvA oscillation measurements which is tuned to efficiently handle neutrino interactions  in the \unit[1--2]{GeV} regime.

In the simulation of the supernova event by the GenieSNova framework, we considered
the visible signal in the \nova detector to be dominated by inverse beta decay
and elastic neutrino scattering on electrons channels.  These two channels
represent $96\%$ of the total expected number of interactions for all interaction channels.
The spectra of the outgoing leptons as a function of time and energy for these channels, as simulated through the GenieSNova simulation chain is shown in figure~\ref{fig:gsnova}.

The carbon de-excitation and other delayed coincidence channels were not
included in the simulation that was used to determine the trigger templates, due to model uncertainties and limitations in the detector readout structure.  However, these channels and other sub-dominant processes,
in the event of a real supernova event, would be fully recorded by the detector
data stream and could be reconstructed in the offline data analysis framework.  These channels may be included in future implementations of the triggering simulations to expand the trigger sensitivities by fully accounting for all expected visible activity in the detector volume.

\subsection{Detector Simulation}

Once GenieSNova has produced a time profile of supernova neutrino
interactions and their daughter particles that would appear in a NOvA
detector, the particles are fed to the same 
detector simulation used in the beam neutrino analyses.  No
modifications to the geometry are needed for the supernova signal and
all standard calibrations are applied to the detector response as are
applied in the beam neutrino oscillation analyses.  This simulation includes the light
yields of the NOvA liquid scintillator using a custom software package developed and tuned to NOvA data and measurements of light yields
made by the NOvA collaboration~\cite{scint_nim}.  This tuning accounts
for both scintillation light and Cherenkov light that is generated
within the liquid and transported through the wavelength-shifting
fiber optics.  The propagated light is then fed to a model for the
response of the NOvA avalanche photodiodes (APD) and a simulation of
the readout electronics, digitization, digital signal processing, and
readout chain.  The results are formatted and stored in the same
format as data from the physical detector systems, which allows the
information to be subsequently used as input to the trigger decision
system.

\subsection{Background data overlays}

The observable signals in both the IBD and elastic scattering channels
have mean energies between 10 and \unit[20]{MeV}, with single detector cell
responses below \unit[10]{MeV}.  This places these signals close to the noise
floor and zero suppression threshold for the digitization and readout
electronics.

In order to accurately model detection algorithms, these signals need
to be placed in the context of the expected cosmic ray induced
activity and the electronics noise that is present on the front-end
electronics.  To accomplish this, ``minimum bias'' data is collected
from the detector.  The minimum bias data stream produces a sequence of complete \unit[5]{ms}
time windows of detector readout.  These \unit[5]{ms} windows are continuous and contiguous in
time ({\it i.e.}, each \unit[5]{ms} window contains all activity for that
time window and sequential \unit[5]{ms} windows in the data stream
used for the supernova analysis are contiguous such that when combined
they provide no temporal gaps in the readout).  The window length of
\unit[5]{ms} is chosen to correspond to the size of the milliblock which is the data
acquisition system's internal discretization of the continuous readout
stream.  This minimum bias data is considered to be free of supernova
induced activity, and simulated supernova signal events are then
overlaid on top of data to produce a data/Monte Carlo hybrid readout stream
which represents the expected full detector response to a supernova event.
This data stream is formatted identically to the actual detector
data, but includes both the simulated signal and real backgrounds.  These
``overlay'' files are then used to tune and test reconstruction and
trigger algorithms and calculate signal selection efficiencies and purities.

\section{Detection of neutrino interaction candidates}
\label{sec:selection}

\begin{figure*} \centering
  \includegr{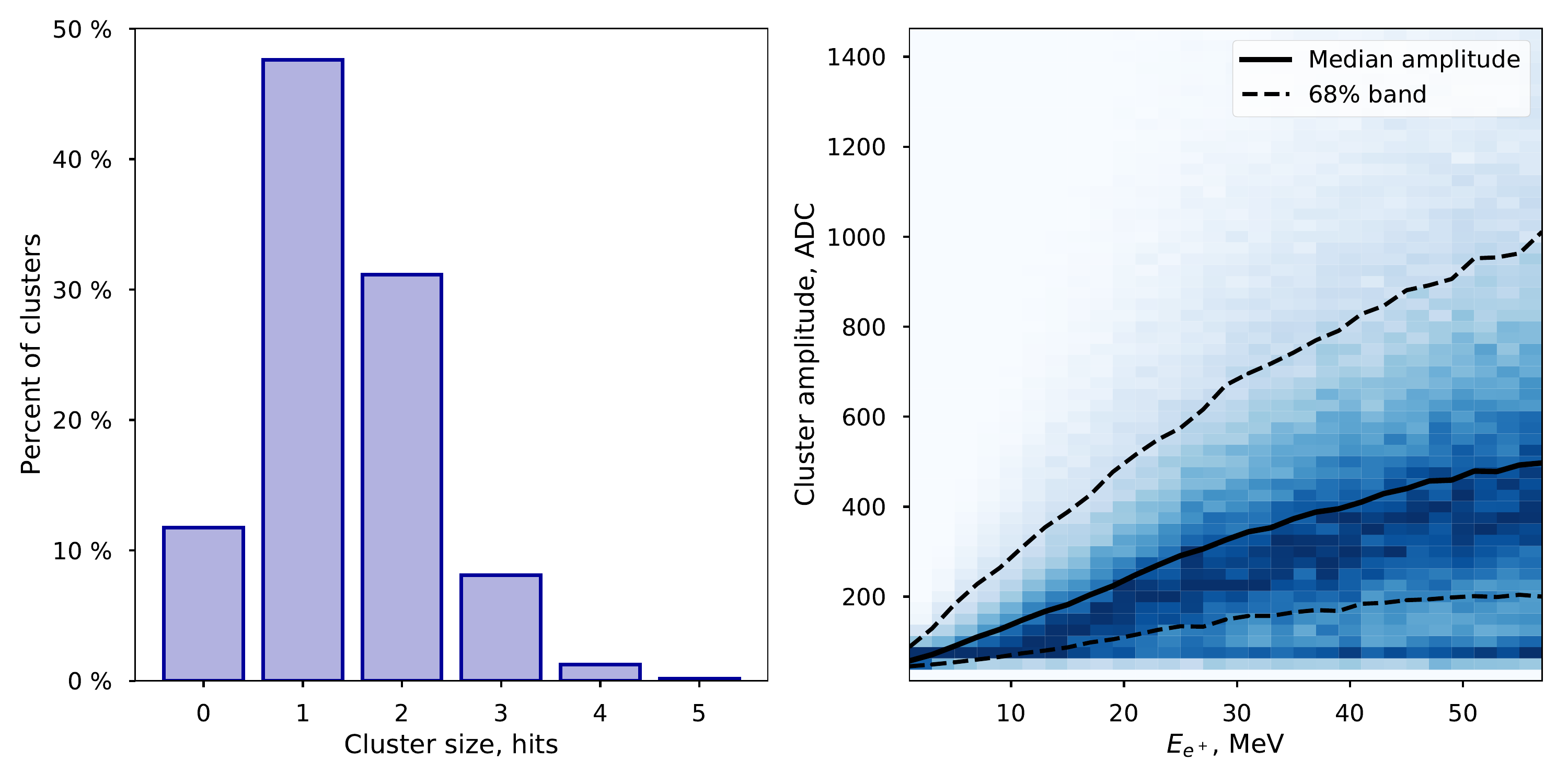}
  \caption{Distribution of the number of hits from simulation, produced by
    positrons from supernova neutrino IBD interactions in the NOvA \fardet using \unit[9.6]$M_\odot$ progenitor model (left), and their summary amplitude in ADC counts
    (right). The solid line shows median amplitude, and dashed lines
    show the 68\% band of ADC vs positron energy.}
\label{fig:positrons}
\end{figure*}

Each neutrino interaction channel under consideration results in an
observable signal  from a low energy electron,
positron, or photon. Identifying these low energy signals is
challenging in both the Near and Far NOvA detectors. Both detectors
are optimized for calorimetry of GeV-scale electromagnetic showers,
corresponding to the electron neutrino appearance topology from the
\unit[2]{GeV} NuMI neutrino beam~\cite{nue_paper}. These showers
extend over distances of meters in the low-Z NOvA detectors and
consist of tens of hits above the zero suppression
cutoff\footnote{The mean number of hits for a muon track arising from
  a $\nu_\mu$ interaction reported in a \nova{} detector is
  122.}, while IBD positrons from a supernova have detectable
energies of \unit[10--30]{MeV}. These particles induce far smaller
signals, with only 1--4 hits
being registered above the zero suppression threshold in the
scintillator cells. This response to low energy particles is shown in
the left of Fig.~\ref{fig:positrons}. In the NOvA detectors, this
response is further suppressed by the attenuation of the scintillation
light through the active medium and along the fiber optic path used
for readout.

Even weaker \unit[8.6]{MeV} photon signals from IBD neutron capture on ${}^{35}\mathrm{Cl}$ will also be present. Given the presence of a large background, such signals are difficult to match with the positron signal, due to their time and space separation. Therefore the IBD neutrons are not considered in this work.

\subsection{Background rejection}

The difficulty of detecting low energy interaction hit clusters is
compounded by the rates of background activity in the NOvA
detectors. In contrast to other supernova neutrino experiments
currently operating~\cite{sn-review}, the NOvA detectors have
considerably smaller overburdens with which to shield the detector
volumes from sources of cosmic induced backgrounds.

The \unit[300]{t} \neardet has the larger overburden of the two
detectors at 225 meters water equivalent (mwe). At this depth,
the average cosmic ray rate results in \unit[37]{Hz} of observed activity in the \neardet. This
rate is sufficiently low that it has minimal impact on the efficiency
for identification of low energy hit clusters. In contrast, the \fardet building is located on the surface, where the overburden is \unit[3.6]{mwe} resulting in
an observed average \unit[148]{kHz} rate of cosmic ray muons,
electromagnetic showers and other cosmogenically induced activity~\cite{nova_numu}. This activity comes from the direct passage of
cosmic rays through the active volume of the detector, as well as from
secondary neutrons, delta rays, spallation products, and Michel
electrons from muon decay.
The first step for a trigger trying to find low energy clusters consistent with neutrino interactions from a supernova is thus to remove hits from other physics processes or readout noise.  The remaining hits can then be examined further.  

\subsubsection{Atmospheric muons}
Michel electrons coming from decay of atmospheric muons with a kinematic upper edge at \unit[53]{MeV} are of comparable
energy to the IBD signal positrons from supernova neutrinos. They form the
primary source of background for the supernova signal, but can be
suppressed through a process that associates a low energy cluster with
the stopping muon track that is the parent of the Michel electron.

In this analysis, muon track trajectories are identified and
reconstructed using a tracking algorithm based on the Hough transform \cite{nova_reco}.
Termination points for these trajectories that occur within the
detectors' fiducial volumes are treated as the endpoint or stopping
position for the muons. Hits associated with the muon track are removed from consideration.  Additionally, hits separated from the track endpoint by less than \unit[32]{cm} and delayed by no more than by \unit[10]{$\mu$s} are considered to be produced by the Michel
electron and rejected from the signal sample for the
neutrino burst signature. 
Note that the same similarity between the spectra of SN induced IBD positrons and Michel electrons which makes them an important background to remove also allows us to benchmark our simulations, as Michel electrons are a standard calibration for \nova.

\subsubsection{High energy showers}
Extensive air showers and other high energy activity induced by cosmic rays, including catastrophic bremsstrahlung interactions, can
produce high hit multiplicities in the NOvA detector.  Many of the
clustering and tracking algorithms used for low energy event
reconstruction are susceptible to failures under these conditions
both in terms of physics content as well as in terms of algorithmic
scaling behavior, as many of the clustering methods have $\mathcal{O}(n_{hits}^2)$ algorithm complexity.  For this reason, high multiplicity time windows
in the data are excluded from the trigger decisions.  Time windows of \unit[1.1]{$\mu$s}
containing hits with a total amplitude exceeding \unit[1,000,000]{ADC} for the \fardet and \unit[300,000]{ADC} for the \neardet are excluded.  In addition, a trailing
exclusion window of \unit[350]{$\mu$s} is imposed to allow the
detector electronics to recover from known response
effects related to high instantaneous rates.  Remnant track segments
are removed after this process by imposing a maximum size requirement
on candidate positron clusters.  This process is discussed in
Sec.~\ref{sec:multiplet}.


\subsubsection{Electronics noise in readout channels}
\label{subsubsec:noisemap}
Zero suppression thresholds for the front-end electronics are set
based on the measured noise envelopes produced by APDs and their
corresponding amplification circuits.  The noise profiles for the
circuits are assumed to be approximately Gaussian, and as such the
zero suppression threshold is set at $+4\sigma$ and $+5\sigma$ from
the channel baseline for the Far and Near detectors respectively.
At these suppression levels, the dark current and leakage current in
the detection circuits produce single hits in the readout that are not
correlated with particle trajectories or interactions in the
detectors.  

The APDs and front-end amplifier/digitization boards used in NOvA are sometimes subject to electrical failures which can cause individual channels to give either excessive or suppressed hit rates.  Channels experiencing these failures are classified as degraded and excluded from the analysis.  These electronics are replaced at regular maintenance intervals, resulting in an average dead fraction of $0.12$\%  for \fardet channels and $2.5\%$ for \neardet channels.  To mitigate the effect of this loss prior to their replacements during maintenance, a map of the average hit rates on a per channel bases is
calculated and used to mask off degraded channels.  For the purposes of the supernova trigger system, channels are masked in the trigger software if they have an inactivity fraction exceeding {90}\% (cold channel) or have a single channel hit rate in excess of \unit[1]{kHz} when averaged over a one hour period (hot channel).  The activity map used to provide this channel masking is recalculated hourly with the maps being archived in the trigger system.  When a channel is tagged for exclusion, it remains excluded from the supernova trigger analysis until it is observed to respond within normal operational parameters for a period of 24 hours.  It is then re-enabled in the subsequent hourly update to the trigger channel masks.

The effect of degraded channels (both hot and cold) on the detection of simulated signal is taken into account by applying the channel masks, obtained from the detector minimum-bias data. Due to the low fraction of excluded channels, the impact on the resulting efficiency of the supernova signal detection is small, compared to other background rejection procedures.

The total integrated rate of single channel noise after the background hits suppression across the 344,064 channels of the \fardet is \unit[56.3]{MHz}.  This rate includes both instrumental noise and low-level light from the environment or sub-threshold particles. Cross-talk between adjacent channels can cause correlations in these rates
which are able to mimic the low energy multi-hit clusters that are
being searched for in the supernova analysis.  However, the X-view and
Y-view readout systems are electrically and spatially decoupled and
there are no known couplings between the readouts of adjacent detector
planes.  As a result single-hit
sources of noise are significantly suppressed by requiring a temporal coincidence between hits
in adjacent planes.  During the part of the clustering algorithm discussed in Sec.~\ref{sec:multiplet}, a temporal coincidence
of less than \unit[32]{ns} is required for a candidate cluster to be formed.  At this
level of matching, the total noise
rate is further suppressed by a factor of 240 from \unit[56.3]{MHz} to \unit[232]{kHz}.

\begin{table*}[!t]
    \centering
    \caption{Average rejection of single hits from various background
      sources in the NOvA detectors: the first row is the total rate, then the effects of each cut follow, with the last row being the remaining rate of single hits.  Low-rate events such as the high energy showers and Near Detector cosmics have a high variance, since very few time windows are affected quite strongly by such activity.}    
\begin{tabular}{llrrr}
\toprule
& {} & Average hit rate, kHz &  Fraction & Variation \\
\midrule
\multirow{6}{*}{\rotatebox[origin=c]{90}{\textbf{Far Detector}}}
& Total detector hit rate               &              74971.09 &   100.00\% &     1.21\% \\\cline{2-5}
& Hits from cosmic ray muons            &              16702.19 &    22.28\% &     5.06\% \\
& Hits from Michel electrons            &               4727.62 &     6.31\% &     5.18\% \\
& Single channel noise                  &               1533.98 &     2.05\% &     2.04\% \\
& High energy cosmic shower activity    &                 96.26 &     0.13\% &   839.35\% \\\cline{2-5}
& Activity after background suppression &              56344.94 &    75.16\% &     0.89\% \\
\midrule
\multirow{6}{*}{\rotatebox[origin=c]{90}{\textbf{Near Detector}}}

& Total detector hit rate               &                715.14 &   100.00\% &    11.01\% \\\cline{2-5}
& Hits from cosmic ray muons            &                  2.57 &     0.36\% &   255.36\% \\
& Hits from Michel electrons            &                  1.14 &     0.16\% &   295.76\% \\
& Single channel noise                  &                445.88 &    61.90\% &    10.60\% \\
& High energy cosmic shower activity    &                  0.04 &     0.01\% &  8796.71\% \\\cline{2-5}
& Activity after background suppression &                269.89 &    37.74\% &     9.03\% \\
\bottomrule
\end{tabular}
    \label{tab:hits}
\end{table*}

The effectiveness of the selection requirements employed to reduce contributions to the single hit rate from
atmospheric muons ($\sim$\unit[100]{kHz} at the Far Detctor, $\sim$\unit[30]{Hz} at the Near), high energy showers ($\sim$\unit[5]{Hz} at the Far Detector), and electronic noise are summarized in Tab.~\ref{tab:hits}.  These are drawn from a typical one minute 
time period of minimum bias data in the detectors, showing the background rates close to the average level.  The milliblock-to-milliblock variation for high-rate backgrounds is quite steady.  Comparatively rare backgrounds which strongly affect few milliblocks have high variance. 

\subsection{Clustering algorithm}
\label{sec:multiplet}
The hits remaining after the removal of background activity, as summarized in Tab.~\ref{tab:hits}, potentially belong to low energy neutrino interactions.  The identification of these interactions requires a multi-stage process of hit clustering and pattern recognition which form the interaction candidates.  The first set of clustering algorithms uses the timing information associated with each hit in the NOvA detector.  Single-hit timing in the NOvA detectors is determined by fitting and interpolating the waveform data that is acquired from the front-end digitizers to the idealized shaper circuit response function of the amplifier circuit.  The process uses a set of four samples separated by \unit[500]{ns} at the Far Detector (\unit[135]{ns} at the Near) to determine the rising edge of the response pulse and the time, $t_0$, of the activity that caused the light.  This process results in a single-hit timing resolution of \unit[8--12 ]{ns}.  This allows for a simple clustering algorithm to be applied based on temporal activity.  Cosmic ray muons are easily identified using this method, as $\beta \approx 1$ particles result in crossing times of $\mathcal{O}$(\unit[100]{ns}) and yield on the order of 400 hits across the detector at near vertical angles\footnote{The NOvA \fardet contains 384 cells per plane.}.  
In contrast, the low electronics noise rates in the detector result in random noise hits, uncorrelated in space and time.
The readout time window is sliced into sub-windows based on these temporal hit distributions.  Spatial clustering is then performed to localize regions of correlated activity.  
\textbf{}
After clustering is performed, each of the resulting activity candidates needs to have additional reconstruction algorithms applied to determine if the topological nature of the activity is consistent with low energy interactions or something else, such as an electromagnetic shower or a muon track.  The standard NOvA event reconstruction algorithms are not well matched to this task due to the span of energies over which the identification is performed, ranging from the multi-GeV regime that dominates the cosmic ray flux seen by the detector, to the GeV-scale beam neutrino interactions for which the detector geometry was optimized, down to the tens of MeV level which characterizes the supernova neutrino energy spectrum.  In particular, low energy interaction candidates are dominated by clusters with  multiplicities insufficient to form a track or shower in the reconstruction algorithms used for oscillation analyses. A typical NuMI beam neutrino interaction produces tens of hits, so these algorithms generally require a minimum of 4--10 hits distributed across many detector planes with no more than a few gaps.  Low energy interaction reconstruction instead relies on hit density and proximity, and does not place a minimum hit requirement on the candidate cluster.  Thus, candidates for low energy interaction reconstruction are taken as those hit clusters remaining after higher energy processes are filtered out.  

The supernova signal IBD interactions are expected to have visible energy signatures in the tens of MeV based on their parent neutrinos' spectrum.  The NOvA detector's cellular geometry and 63\% active fraction result in an approximate energy deposition of \unit[8--10]{MeV} per cell crossed by the IBD positron.  At these expected energy deposition rates, the upper tail of the neutrino energy spectrum is expected to produce hit clusters which extend between 1--4 cells (\unit[6--26]{cm}) corresponding to energies up to \unit[40]{MeV}.  Individual channels have a lower threshold for hit detection that is tuned to suppress noise from the electronics.  This suppression threshold corresponds to the equivalent light output of approximately 71\% of the average response of a minimum ionizing particle passing through the far end of a detection cell.  This results in a worst-case detection threshold of approximately \unit[8]{MeV} for interactions near the bottom and east edges of the detector, at the opposite ends of the cells from where the front-end readout boards are located.

The neutrino interactions expected from a supernova thus produce very small clusters containing only a few hits.  Any such clusters of hits are potential neutrino interaction candidates, and are reconstructed as described in the following sections.

\subsubsection{Position selection}
\begin{figure}
    \centering
    \includegraphics[height=6cm]{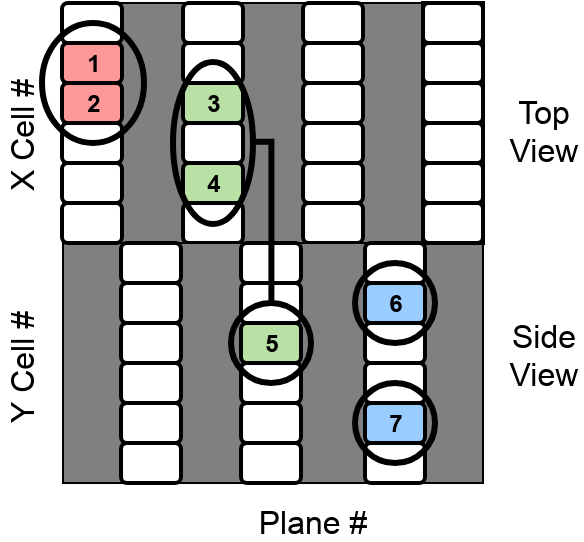}
    \caption{Illustration of cluster finding procedure for 7 example hits (numbered). Hits 1 and 2 will form a cluster only in the X-view. Hits 3, 4, and 5 will form a cluster in both X- and Y- views (this cluster can be considered a neutrino interaction candidate). Hits 6 and 7 will not be connected to any group, as they are too far apart.  The Z-axis is left-to-right in this figure.}
    \label{fig:clustering}
\end{figure}

Hits belonging to the same cluster must be separated by not more than one cell within the same plane and contain cells from neighboring scintillator planes, meaning that they must contain cells of different orientation (X and Y).  Clusters with only hits in cells on the same plane (X and X, or Y and Y) are not considered in this analysis, as lack of a 3D position prevents the next step (time correction). Figure~\ref{fig:clustering} illustrates the selection of the clustering algorithm.

\subsubsection{Time correction}
\begin{figure*}[!t]
\centerline{
\includegraphics[width=0.9\linewidth]{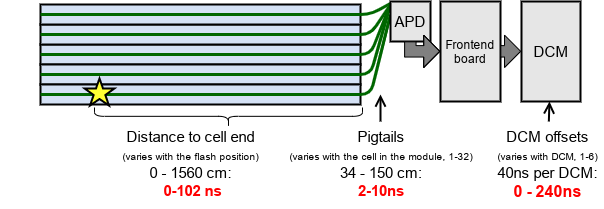}
}
\caption{Timing corrections applied to the scintillation hit: starting with the light propagation time through the cells, then through optical fiber pigtails, into a unit of the front-end electronics and finally to the Data Concentrator Module (``DCM'').}
\label{fig:timing}
\end{figure*}

The \nova{} detectors' front-end readout systems are fully synchronized and have corrections applied at the hardware level for signal propagation delays within the timing and clock systems.  However, these corrections do not include corrections for the signal propagation between the daisy-chained timing links that connect the Data Concentrator Modules (DCMs) that are physically positioned on the top and side faces of the detectors.  These propagation delays have been measured for each of the DCM positions, and range from \unit[15--120]{ns} corresponding to the near and far sides of the readout chain.  These timing corrections are applied in the software layer of the supernova trigger to the raw timestamps read out for each of the hits in the detector in order to properly match scintillation signals produced in different cells across the detector.  
These corrections are the known time shifts in the detector synchronization chain~\cite{timingProceeding} and the propagation delay resulting from the scintillation light being transported along the wavelength shifting fiber optic path to the APD readout at the end of the detection cell (Fig.~\ref{fig:timing}).
The correction algorithm takes into account the delay for the light transportation in the fibers, using the fiber length and measured index of refraction.  The algorithm assumes an origination point for the light to correspond to the intersection of the X and Y hits in the candidate cluster, and computes the distance to the corresponding APD readouts for X-view and Y-view planes. The application of this correction restricts the analysis to considering only clusters containing both X and Y hits.  Since X-Y pairs are already required to reduce the accidentals rate from correlated noise in the electronics, this correction does not result in an additional degradation of the signal detection efficiency.

The single-hit timing resolution of the \nova{} near and far detectors has been measured both in situ using cosmic ray muon tracks and in a dedicated teststand which was used to precisely characterize the front-end amplifier and shaper response functions that are needed for the calculation.  These measurements result in a single-hit timing resolution function which depends on the photo response of the cell.  This results in an expected timing resolution of \unit[12]{ns} at a \unit[200]{ADC} signal level, and less than \unit[8]{ns} above \unit[400]{ADC}.  This resolution is used to set the timing coincidence level for the hit matching at \unit[32]{ns}, corresponding to an envelope allowing for approximate fluctuations of \unit[2.0]{$\sigma$} for the lowest light response levels.

\subsection{Candidate selection}
\label{sec:candSelection}
The siting of the \nova{} \fardet on the surface under a \unit[3.6]{mwe} overburden results in a significant flux of cosmic ray induced activity in the active volume of the detector.  In particular, this flux includes electromagnetic shower activity which originates from interactions in the overburden of the detector hall and can mimic the low energy interaction topology.  To reject this background source, a fiducial volume cut is applied to the data.  This cut  rejects activity coming from particle and shower penetration into the detector volume.  The fiducial region has a boundary of 16 cell widths (\unit[62]{cm}), corresponding to 1.5 radiation lengths in both the X-view and Y-view from the bottom and sides of the detector.  The boundary is set to 24 cell widths (\unit[94]{cm}), corresponding to 2.4 radiation lengths from the top face of the detector to provide additional attenuation of electromagnetic activity and to reduce the contamination of the signal from neutrals penetrating deep into the detector prior to interaction. For the Near Detector, the cut is 8 cells for X and Y and 8 planes for Z.

\begin{figure*}[!t]
\centerline{
\includegr{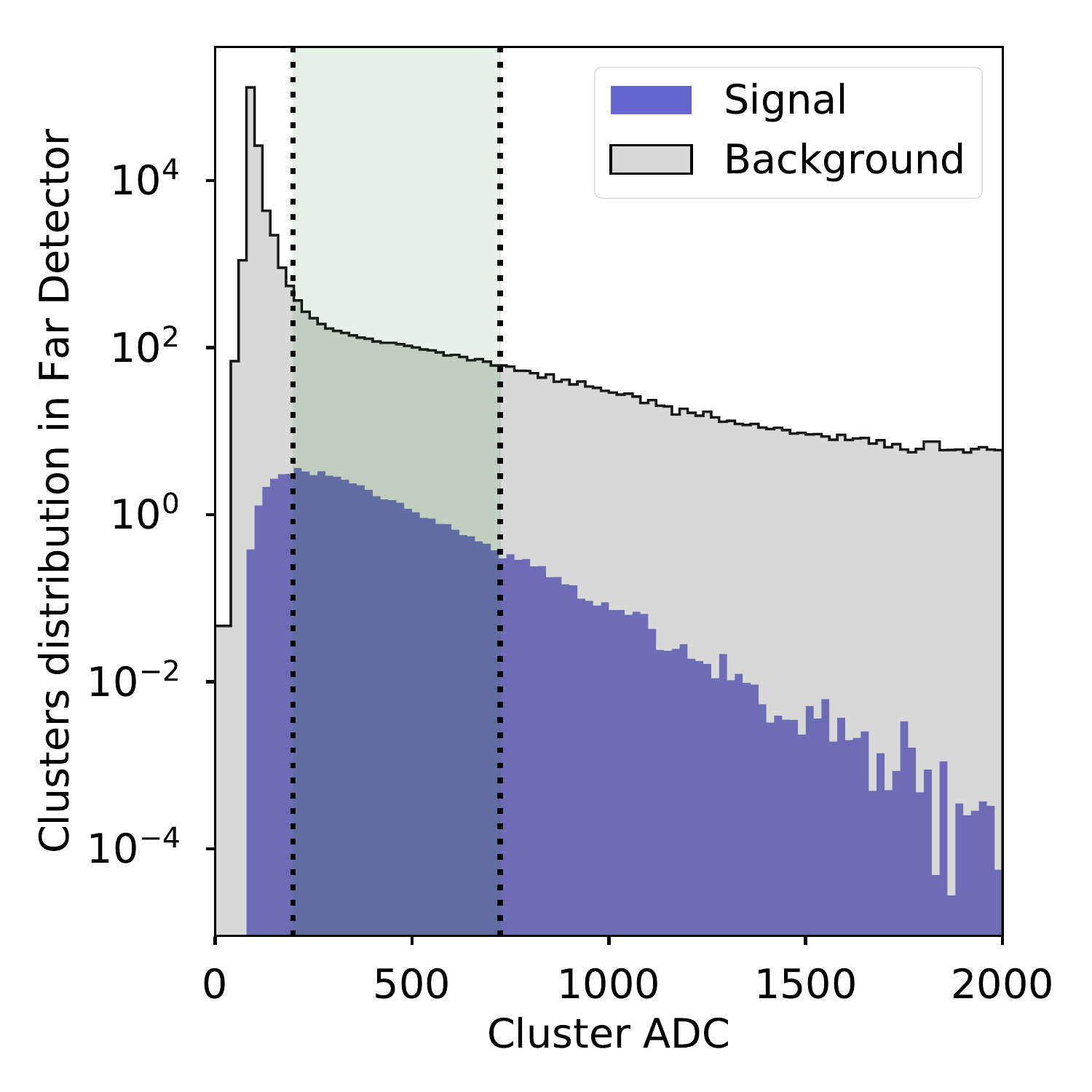}
\includegr{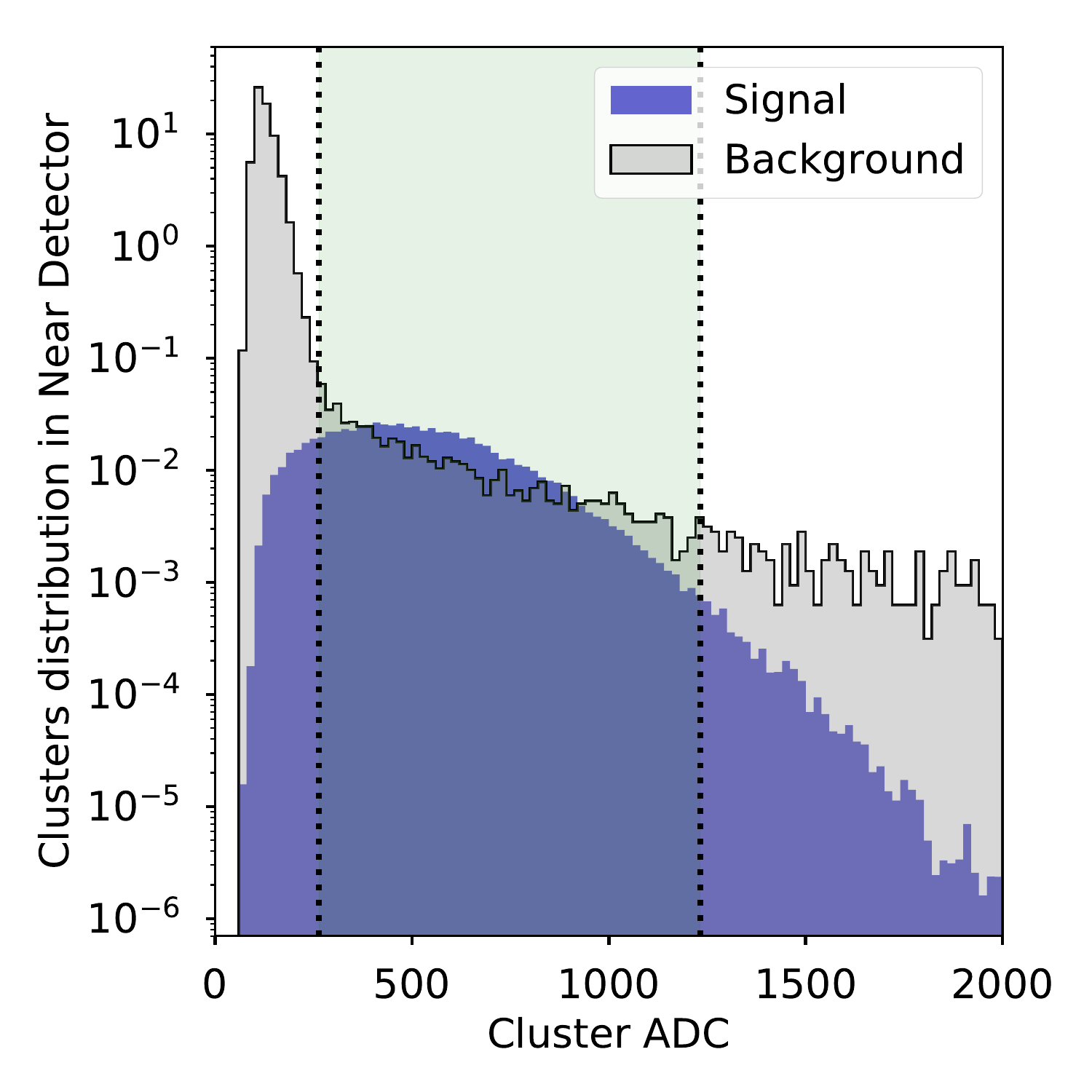}
}
\caption{Selected ADC range for candidate clusters for the \fardet (left) and \neardet (right).  The signal clusters are from the Monte Carlo simulation of $\unit[9.6]{M_\odot}$ supernova at \unit[10]{kpc} distance and the background clusters are from minimum bias data.  The range between the vertical dotted lines shows the ADC cut (from Tab~\ref{tab:selection}) used to select the best $N_{sg}/\sqrt{N_{bg}}$ ratio for inclusion in the time series shown  in Fig.~\ref{fig:time_correlation}.}
\label{fig:adc_selection}
\end{figure*}
Finally, we apply a cut on the total ADC value of all hits in the cluster (Fig.~\ref{fig:adc_selection}).
A range of acceptable ADC values was chosen to maximize the ratio of signal to the average background fluctuation {$N_{sg}/\sqrt{N_{bg}}$}, eliminating low energy noise hits and higher energy cosmogenic remnants.
Table \ref{tab:selection} summarizes the ADC and fiducial volume cuts applied to each cluster to be considered a neutrino candidate.
\begin{table}[!t]
    \centering
    \caption{The selection criteria cuts for neutrino interaction candidates which were tuned as shown in Fig.~\ref{fig:adc_selection} and Sec.~\ref{sec:candSelection}.}
    \begin{tabular}{ccc}
        \toprule
        Cut & \neardet & \fardet \\
        \midrule
        ADC range & $[280, 1430]$ & $[230 , 910]$ \medskip\\
        \multirow{3}{*}{Fiducial volume} 
            & $8\leqslant \text{X cell} \leqslant 88$ & $16\leqslant \text{X cell} \leqslant 368$\\
            & $8 \leqslant \text{Y cell} \leqslant 88$ & $16 \leqslant \text{Y cell} \leqslant 360$\\
            & $8 \leqslant \text{Z plane} \leqslant 184$& $8 \leqslant \text{Z plane} \leqslant 888$\\
        \bottomrule
    \end{tabular}
    \label{tab:selection}
\end{table}

\subsection{Removing time-correlated candidate groups}
\label{sec:grp_remove}
\begin{figure*}[!t]
    \centerline{
    \includegr{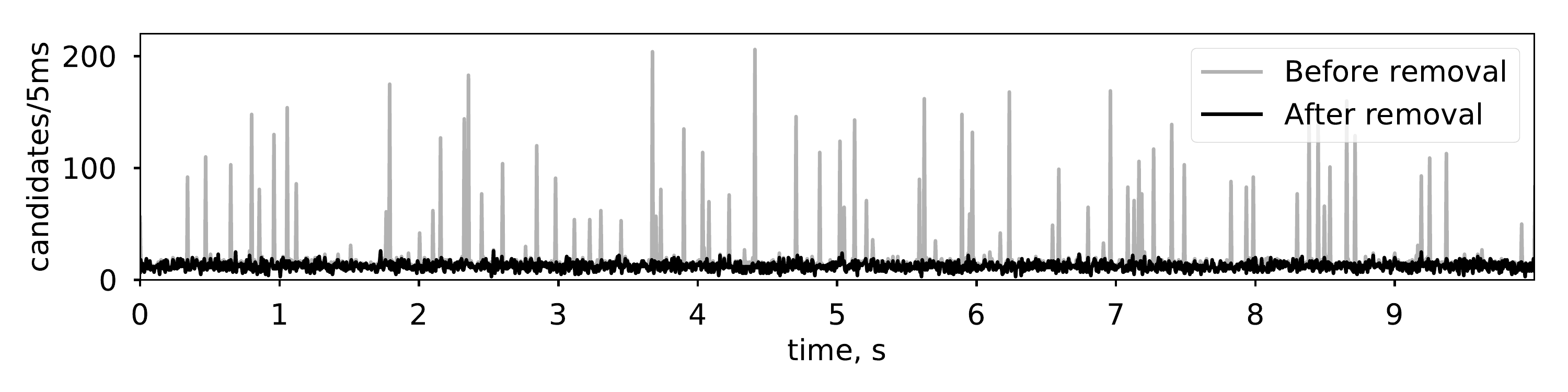}
    }
    \caption{A time series of potential SN neutrino candidates seen at the \fardet.  Supernova neutrinos would be scattered randomly in time while remnants of electromagnetic interactions are correlated in time and appear in the form of peaks (grey).  This background can easily be removed by cutting out the peaks (see Sec.~\ref{sec:grp_remove} for details), causing 0.15\% deadtime.}
    \label{fig:time_correlation}
\end{figure*}

After the background rejection procedures described above, there remains some rate of non-neutrino candidates produced by electromagnetic showers induced by the cosmic rays in the detector.
The neutrino interactions from supernovae should not be correlated on short timescales, so we reject any pair of interaction candidates with timestamps closer than \unit[250]{ns}. 
Since the number of candidates at this step is low, this rejection produces a dead time less than $0.15\%$. This rejection significantly decreases the variation of the background candidates in time (Fig.~\ref{fig:time_correlation}) so that the background level follows a Poisson distribution.

\subsection{Selection results}

\begin{figure*}[!t]
\centerline{
\includegr{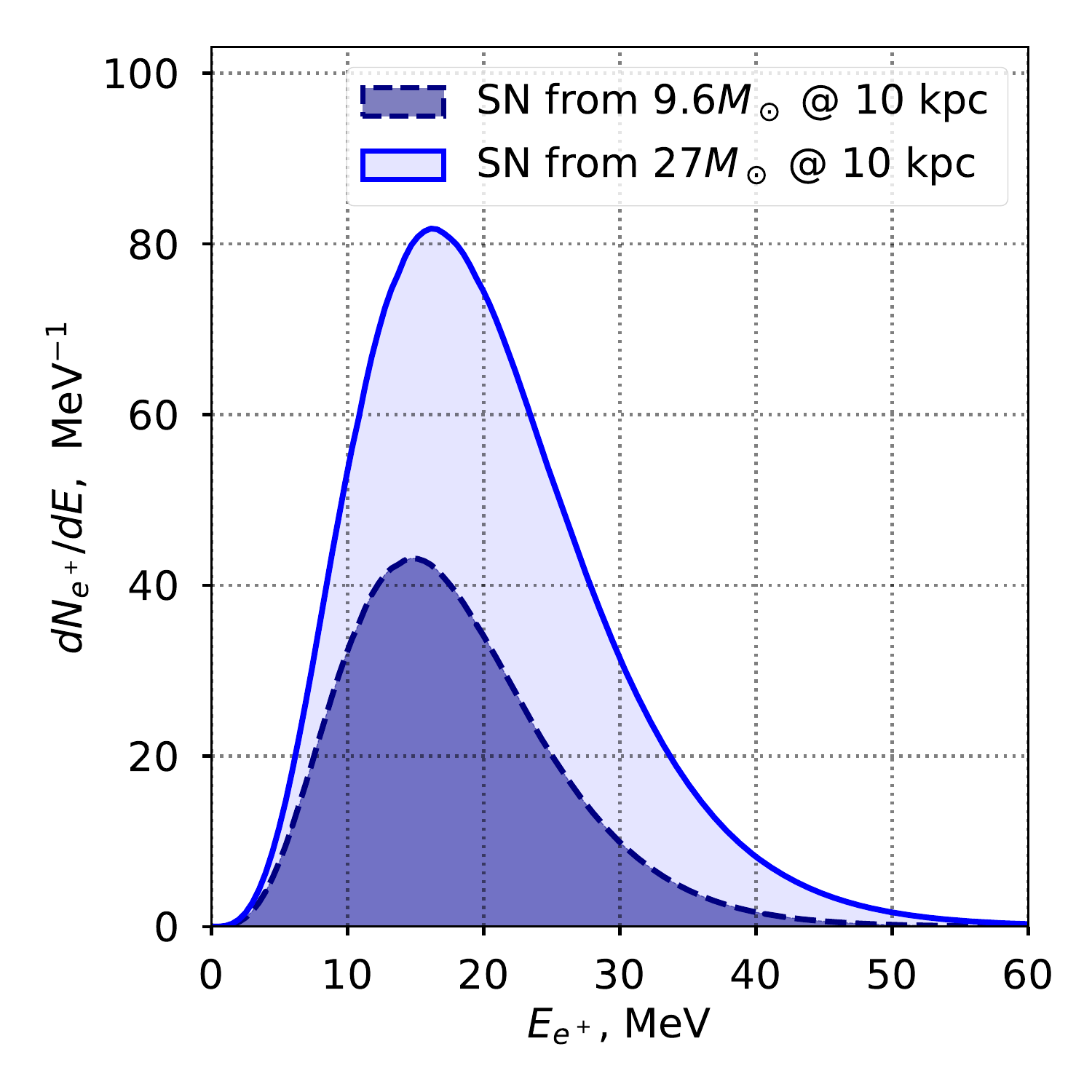}
\includegr{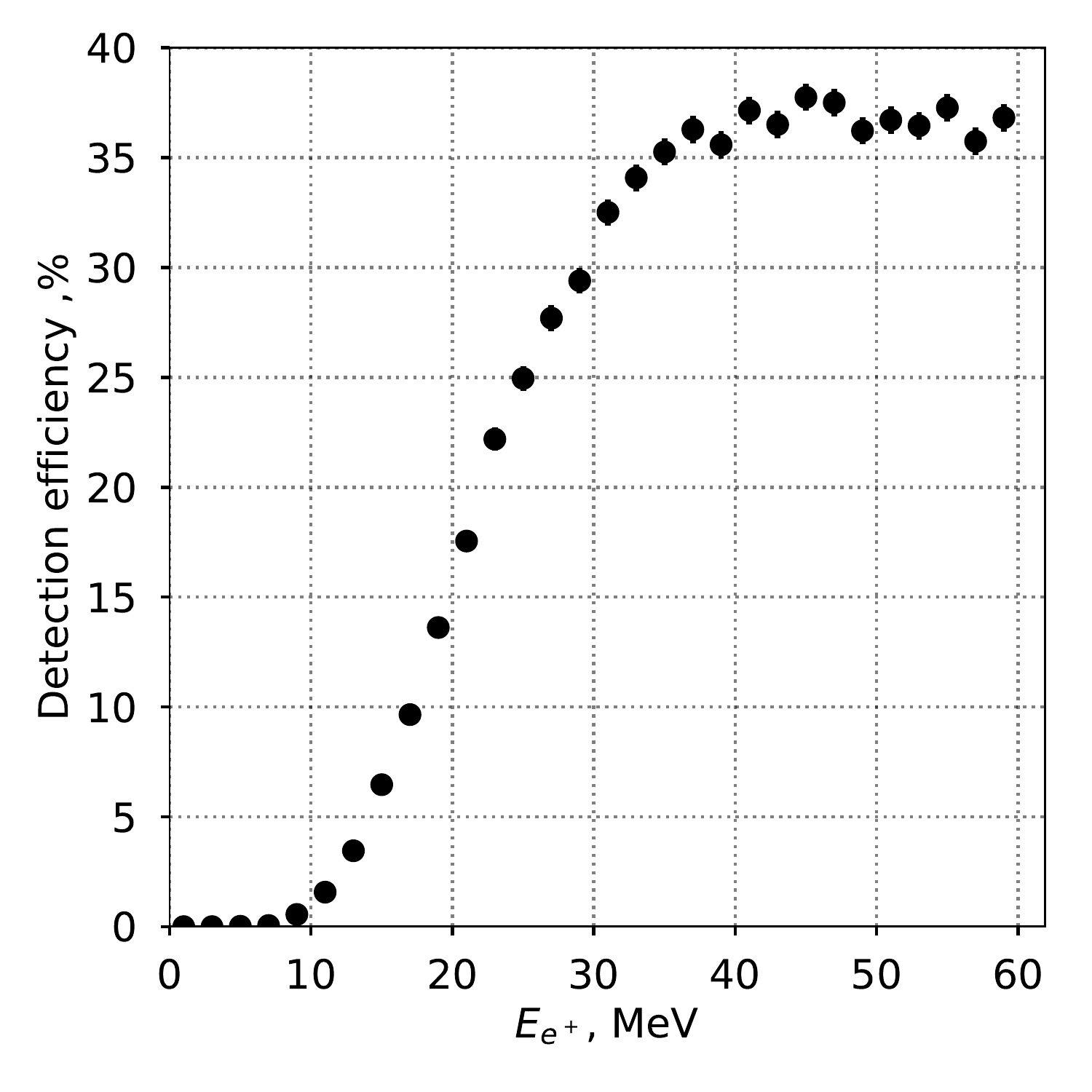}
}
\caption{Positron spectrum from simulated supernova neutrino IBD interactions in the \fardet (left) and efficiency of positron detection as a function of positron energy (right) with the cuts summarized in Tab.~\ref{tab:sel_cands_FD}.}
\label{fig:posit_efficiency}
\end{figure*}

Applying the described selection procedure, we can estimate the detection efficiency using a simulated sample of positrons uniformly distributed within the detector and with random directions.  This allows the calculation of 
the efficiency of detecting an individual positron as a function of positron energy, as shown in Fig.~\ref{fig:posit_efficiency}.
This produces a signal to noise ratio of {1:29} at the FD and {2.5:1} at the ND for a simulated \unit[9.6]{M$_\odot$} supernova at \unit[10]{kpc}.
Table~\ref{tab:sel_cands_FD} summarizes the candidate selection cuts and their effect on the number of candidates during the first second of a supernova.
\begin{table*}[!t]
    \centering
        \caption{The absolute survival efficiencies for each selection cut
      that is applied in the trigger algorithms.  Survival fractions
      are computed independently for each cut as applied to both
      background activity in the detector and neutrino signal corresponding to the first second of a
      \unit[9.6]{M$_\odot$} supernova at \unit[10]{kpc}.  The aggregate
      efficiency of all cuts is also computed.   }
    \begin{tabular}{llrr@{\hskip 25pt}rr}
\toprule
&
\multirow{2}{*}{Cut} &
\multicolumn{2}{c}{Background}&
\multicolumn{2}{c}{Signal}\\
\cmidrule{3-4}
\cmidrule{5-6}
        &       & $N_{bg}$/s & $ \varepsilon$ & $N_{sg}$/s & $ \varepsilon$ \\
\midrule
\multirow{6}{*}{\rotatebox[origin=c]{90}{\textbf{Far Detector}}}
        & Reconstructed clusters &  322811.99 &       100.00\% &     316.24 &       100.00\% \\
        & X and Y hits &  231866.53 &        71.83\% &     145.16 &        45.90\% \\
        & $N_{hits}\leqslant4$ &  310010.78 &        96.03\% &     315.06 &        99.63\% \\
        & Fiducial Volume &  172281.67 &        53.37\% &     118.45 &        37.46\% \\
        & ADC cut &   25879.67 &         8.02\% &     216.38 &        68.42\% \\
        & Total &    2483.21 &         0.77\% &      86.64 &        27.40\% \\
\midrule
\multirow{6}{*}{\rotatebox[origin=c]{90}{\textbf{Near Detector}}}
        & Reconstructed clusters &     403.95 &       100.00\% &       3.16 &       100.00\% \\
        & X and Y hits &     215.64 &        53.38\% &       2.19 &        69.35\% \\
        & $N_{hits}\leqslant4$ &     394.81 &        97.74\% &       3.15 &        99.67\% \\
        & Fiducial Volume &      68.10 &        16.86\% &       1.49 &        47.23\% \\
        & ADC cut &      24.30 &         6.02\% &       2.73 &        86.29\% \\
        & Total &       0.52 &         0.13\% &       1.28 &        40.43\% \\
\bottomrule
\end{tabular}

    \label{tab:sel_cands_FD}
\end{table*}

At this stage, most of the obvious non-SN related candidates have been removed.  The count of remaining candidate neutrinos is passed to the time series discussed in Sec.~\ref{sec:sensitiv}.

\section{Trigger system for supernova detection}
\label{sec:ddt}
\subsection{Data-driven triggering in NOvA}

NOvA's Data-Driven Trigger (DDT) system, described in Sec.~\ref{sec:detectors}, is used to perform fast data reconstruction online in order to decide which time ranges of data should be saved for further offline analyses.  
To achieve this, the data stream from the detector is sliced into \unit[5]{ms} chunks (called ``milliblocks'') and, for supernova detection, data in these milliblocks are subjected to the selection criteria described in Sec.~\ref{sec:selection} to count the number of potential supernova neutrino candidates in each \unit[5]{ms}.

\subsection{Supernova trigger infrastructure}\label{sec:trigger_infrastructure}

In contrast to other NOvA triggers for which the trigger decision is made based on one \unit[5]{ms} milliblock of data, the neutrino signal from a nearby supernova would last for tens of seconds.  It would also likely not be evident in any single milliblock, given the comparatively low signal to noise ratio seen in Fig.~\ref{fig:adc_selection}.  Thus, the accumulated information from hundreds of milliblocks must be used to decide if a supernova has been observed by the NOvA detectors.  This must happen in parallel, as consecutive milliblocks of time are stored on different physical buffer node computers, with the resulting statistics being reported back to a central process on the ``global trigger'' node of the DAQ.


Thirteen processes on any given buffer node computer are running an instance of the analysis chain described above on a milliblock of data resident in the computer’s memory, using eleven of the node's sixteen cores.  
This results in a count of potential positron candidates in that block of data, which the process reports to the central global trigger manager node.  This node sorts these counts into a consecutive time series, in the same way all the other milliblocks of data are counted by many such processes through the DAQ server farm.
This time sequence is convolved with the filtering kernel described in Sec.~\ref{sec:sensitiv} to suppress background fluctuations and enhance the statistical significance of any potential signal.
Should data be found that are consistent with a burst of positrons from IBD interactions of supernova electron anti-neutrinos, the global trigger issues a trigger request to the system, and the relevant time block is saved to disk.  A total of \unit[45]{s} of data is saved, starting \unit[5]{s} before the triggered time to record a baseline of data before the neutrinos arrived.

\subsection{Triggering delay}
One of the main characteristics of the triggering system is a reaction time:
the delay between the time when the neutrino signal from the supernova explosion reaches the detector and the moment when the trigger message would be issued.  When the data is being buffered, too long a delay could result in a trigger issued after the data to be saved have fallen out of the buffer.  Additionally, it is desirable to issue as prompt an alert to SNEWS as possible.  This reaction time is defined by the individual delays in the supernova trigger pipeline. 

We describe the sources of these delays and their average values for the \fardet.
First, the DAQ system must read out detector channels, build the milliblock data slice, and deploy it to one of the $\sim$2200 parallel DDT processes, which takes \unit[3.5]{s} on average.
The DDT processes perform the background subtraction and reconstruction of neutrino interaction candidates as described above. This processing takes about \unit[5]{s} for one milliblock.

Before sending the calculated neutrino candidate rate, the buffer node accumulates ten measurements to form a packed message to send. This is done to reduce the load on the network traffic between the buffer nodes and the global trigger node. Accumulating ten milliblocks on one buffer node takes, on average, \unit[8.5]{s}.

The messages with candidate rates and milliblock timestamps arrive to the global trigger from parallel DDT processes unsorted in time, so additional delay arises from buffering incoming data until a continuous one second of data is populated. This delay varies depending on the time spread of incoming messages. An average value for this delay is \unit[18]{s}.

Finally, the global trigger needs to process at least \unit[5]{s} of continuous data to decide if it contains a supernova signal.

In case any data point is lost (\emph{e.g.} a milliblock was not processed or data was corrupt), the system waits for its arrival for up to \unit[60]{s} before skipping this point in the processing. 
This defines the maximal possible delay.

In the case of the Near Detector, most of the delays for processing and waiting for the data are negligible, because the data flow is much smaller and so the number of parallel processes is reduced to 169.

The trigger system delay was measured as the difference between the milliblock timestamp and the time when it reaches the end of the system pipeline. The total delay for the \fardet is \unit[40]{s} on average and \unit[60]{s} at maximum. The delay for the Near Detector is very stable at the level of \unit[5.7]{s}.  These latencies are sufficiently low to provide a useful supernova trigger: both are much less than the buffer depth, and are among the lowest of any of the SNEWS experiments.

\subsection{Cross-detector triggering}\label{sec:cross_trigger}
NOvA is composed of two detectors both of which are sensitive to SN neutrinos. Once one of the detectors identifies a potential supernova burst, a trigger is sent to the other detector to ensure that data is also saved there.  This cross trigger results in a logical ``OR'' of the individual triggers of the NOvA near and far detectors when calculating the trigger sensitivities.

Future work is planned to combine the time series acquired at each detector.  Combining the two series will result in better rejection of fluctuations and more sensitivity to a real signal.

\subsection{External supernova trigger from SNEWS}
To increase the chance that neutrinos from a galactic supernova would not be missed, the NOvA detectors will also save data based on an external trigger from the SNEWS coincidence network~\cite{snews}.   
Should a coincidence between the supernova triggers of other neutrino experiments be made by the SNEWS project, an XML-RPC message is sent by the SNEWS server to the NOvA DAQ,
triggering a \unit[45]{s} readout of both NOvA detectors’ data~\cite{nova_snews} starting \unit[5]{s} before the SNEWS coincidence time.  This connection is tested by saving a short (\unit[50]{$\mu$s}) time window of data once per minute, and one full-length \unit[45]{s} time block once per day at each detector.  

To help reduce data processing bottlenecks at the event building stage in NOvA, each unit of data will be included in one supernova-related trigger at a time, rather than being included in multiple simultaneous triggers.  For example, a real supernova would ideally produce self-triggers at each detector, as well as cross-triggers from each detector, and an alert from SNEWS.  Any overlapping times from all these long readout windows will be saved only once per detector.  

In the future, in addition to NOvA saving this data, these triggers will be reported to SNEWS to help form a prompt coincidence with other neutrino detectors around the world.

\section{Trigger sensitivity}
\label{sec:sensitiv}

After the parallel DDT processes have performed searches for IBD interaction candidates, data on the candidate rate per \unit[5]{ms} are sorted and accumulated in a time series $\{t_i, n_i\}$.
In order to decide if this time series contains a signal from a supernova, we scan the incoming data and test the ``background-only'' ($H_0 = B$) and ``signal + background'' ($H_1 = B + S$)  hypotheses. These hypotheses predict a Poisson distribution for the number of candidates in the $i$-th time bin, with mean values $\lambda$:

\begin{align*}
P(n_i|H_0)& = \mathrm{Poisson}(n_i | \lambda = B),\\
P(n_i|H_1)& = \mathrm{Poisson}(n_i | \lambda = B + S_i).
\end{align*}

\noindent In order to discriminate these hypotheses based on the data in a time window $\vec{n} = \{n_i\}$, we construct the log likelihood ratio value
\begin{equation}
\label{eq:llr}
\ell(\vec{n}) \equiv \log\prod_i \frac{P(n_i|H_1)}{P(n_i|H_0)} = \sum_i n_i\cdot \log(1+S_i/B)
\end{equation}
and use it as a test statistic. Thus, our discrimination power depends on the average background level $B$ and an expected signal profile $\vec{S} = \{S_i\}$.
Such discrimination is optimal in case the measured signal $\vec{S'}$ coincides with the expected one $\vec{S}$. However, choosing $\vec{S}$ such that it follows the general features of the SN signal also gives results that do not drastically change sensitivity for two very different signal models (see section \ref{ssec:results}).

In order to decide if the observed signal contains evidence of a supernova neutrino burst, we calculate the p-value
$$
p(\vec{n}) = P(\ell>\ell(\vec{n})|H_0),
$$
and, finally, compute the signal significance score in ``sigmas'':
\begin{equation}
    z(\vec{n}) = \mathrm{erf}^{-1} \left(1-2p(\vec{n})\right).
\end{equation}

\noindent The significance score $z$ for the ``background-only'' hypothesis $H_0$, follows standard normal distribution by construction:
$$
P(z | H_0) = \mathcal{N}(z | \mu=0, \sigma=1).
$$

\noindent If the significance score exceeds threshold $z_0 = \unit[5.645]{\sigma}$, the trigger signal is sent. This threshold corresponds to the average rate of one false trigger from background fluctuations per week, the target rate for SNEWS input.

During the trigger system operation the background level $B$ is estimated at the end of every ten minutes period based on the activity in this period. Thus the triggering system can adapt to slow changes in background conditions, maintaining the same false alarm probability: this does not appreciably change the supernova sensitivity. Sudden background changes on timescales less than ten minutes can cause the false trigger alarms, as observed during the commissioning (see section \ref{sec:commissioning}).  Large background changes are indicative of detector problems, and quickly fixed lest all of NOvA's analyses be degraded.
\subsection{Results}
\label{ssec:results}
\begin{figure*}[!htb]
\centerline{
\includegr[scale=0.5]{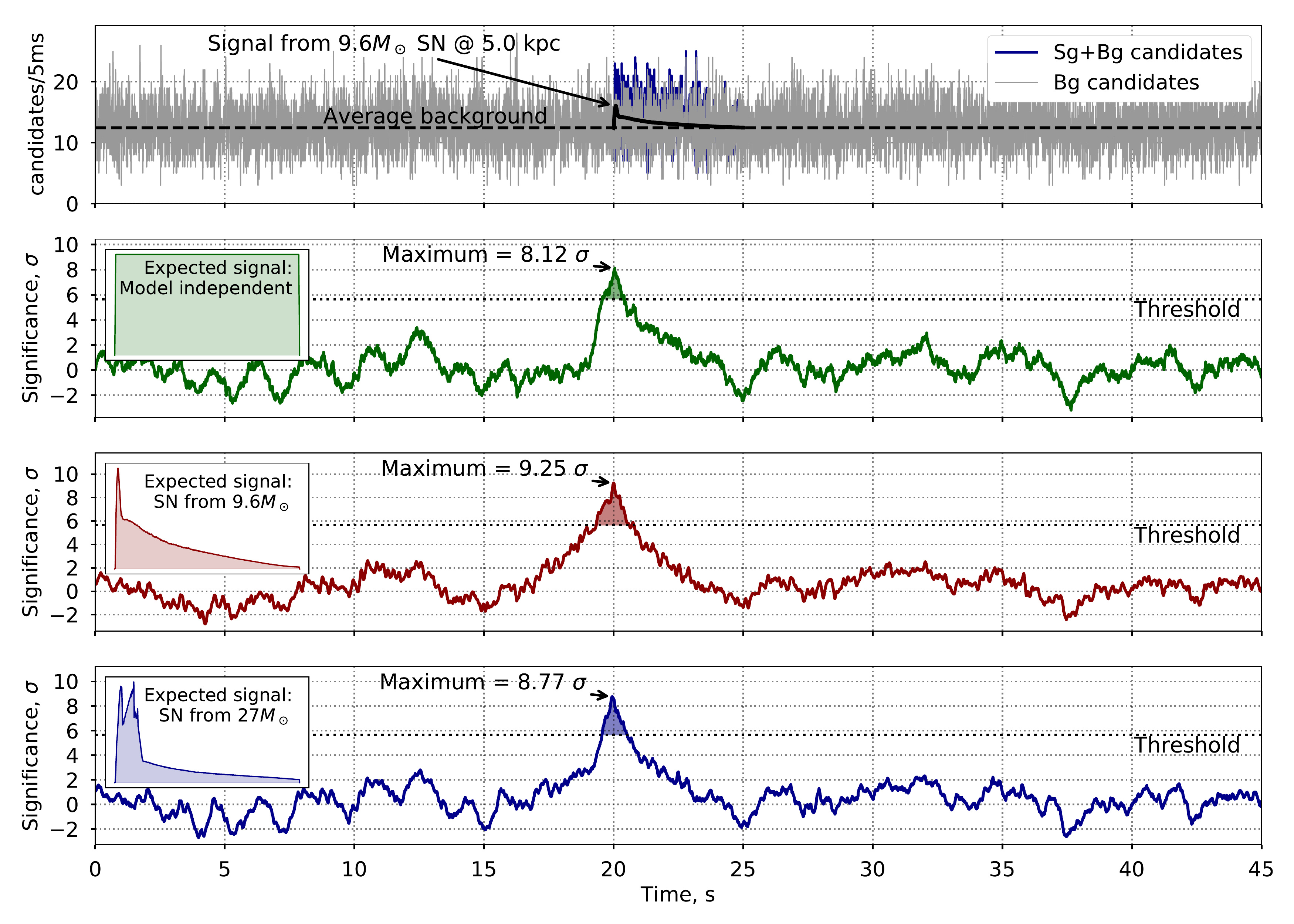}
}
\caption{Example of the trigger system detecting a simulated supernova signal from a star \unit[5]{kpc} away.  The top time series of neutrino candidates per \unit[5]{ms} bin in the NOvA \fardet shows the location and size of an injected signal. The next three time series are the significances of the data matching the expected signal shape  starting at that point in the time series, expressed as Gaussian sigmas, calculated as described in Sec.~\ref{sec:sensitiv}.  The three signal shapes are: a uniform one second increase in rate (top pane); the detector response for a \unit[9.6]{M$_\odot$} progenitor core-collapse model (middle pane); and for a \unit[27]{M$_\odot$} progenitor at the bottom. The injected signal is drawn from the middle model, causing it to match the best: but the other two templates still pick up the correct time ($t=\unit[20]{s}$) with high significance. Models are from~\cite{garching}.}
\label{fig:time_series}
\end{figure*}

Figure~\ref{fig:time_series} shows an example of how the trigger system processes the input data time series of candidates per \unit[5]{ms} time bin ($n_i$) where the counts in the top plot are potential neutrino candidate clusters drawn from actual minimum bias detector data (grey) with the addition of a simulated supernova signal (blue) at $t=\unit[20]{s}$.  The next three plots are time series of the resulting significance scores $z(\vec{n})$ calculated using three types of expected signal shapes: a naive ``top hat'' shape; and two shapes drawn from models~\cite{garching} of specific low-mass (\unit[9.6]{M$_\odot$}) and high-mass (\unit[27]{M$_\odot$} ) progenitor stars.  The shape which was drawn from the model which was injected in this test (the \unit[9.6]{M$_\odot$} model) has the highest significance as expected.  However, both other models also match well enough to produce a SN trigger readout at the correct time.  

\begin{figure*}[!htb]
\centerline{
\includegr[scale=0.54]{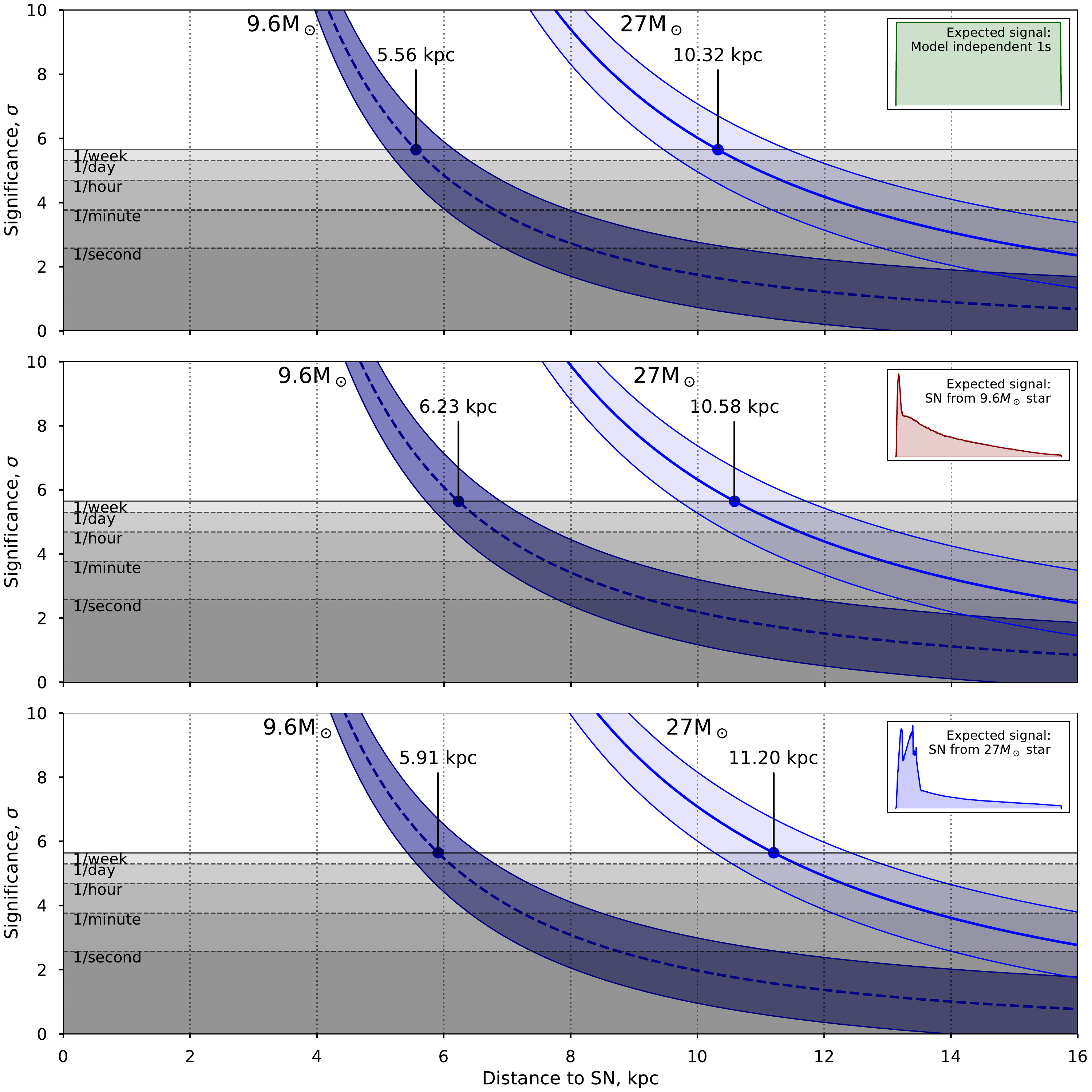}
}
\caption{The expected significance of the supernova signal observations vs.\ the distance to the supernova for \unit[9.6]{M$_\odot$} and \unit[27]{M$_\odot$} progenitor stars in the \fardet, compared to the rate of background induced triggers at various time scales (grey horizontal levels). 
The middle lines show the mean significance value, while the bands show the region with 68\% significance distribution. 
The point at which the mean signal significance is equivalent to an accidental rate of once per week is noted in the plot, in kpc.  The three signal shape templates used, from top to bottom, correspond to those shown in Fig.~\ref{fig:time_series}. 
}
\label{fig:sign_fd}
\end{figure*}

\begin{figure*}[!htb]
\centering
\includegr[scale=0.54]{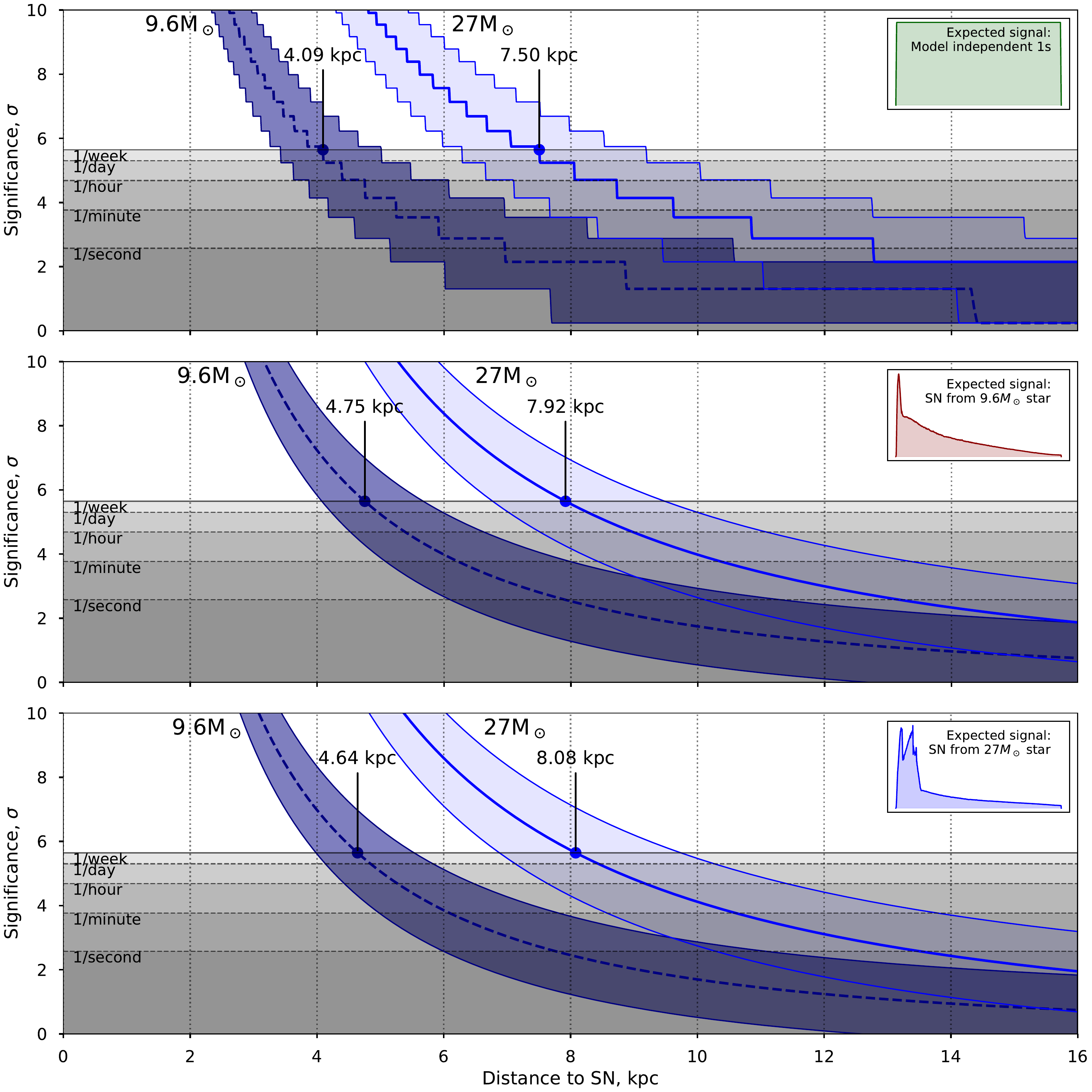}
\caption{The expected significance of the supernova signal observations vs.\ the distance to the supernova for \unit[9.6]{M$_\odot$} and \unit[27]{M$_\odot$} progenitor stars in the \neardet, compared to the rate of background induced triggers at various time scales (grey horizontal levels).  The point at which the mean signal significance is equivalent to an accidental rate of once per week is noted in the plot, in kpc.  The three signal shape templates used, from top to bottom, correspond to those shown in Fig.~\ref{fig:time_series}. The rate at the \neardet is small, leading to discrete jumps in the top figure.}
\label{fig:sign_nd}
\end{figure*}

To decide which shape template to use in the online trigger, the two supernova models were tested against each shape template for both detectors as a function of distance.  Figures~\ref{fig:sign_fd} and~\ref{fig:sign_nd} show the signal significance for both supernova models vs.\ distance for each of the three shapes.  The horizontal bands are significances corresponding to various ``false alarm rates''.  The target rate of accidental triggers for SNEWS input is once per week.  When the threshold is set at that level the intersection of the model's central and that threshold yields the distance sensitivity to a supernova exploding as per that model when searched for using that shape template.   This distance is listed on each plot.

The operational trigger system uses the expected signal shape from the $\unit[9.6]{M_\odot}$ model to make its decision, as it represents the most general features of the supernova neutrino signal. The choice of optimizing for the the dimmer (and likely more commonly occurring) $\unit[9.6]{M_\odot}$ model is validated by the good performance obtained for both the $\unit[9.6]{M_\odot}$ and the brighter $\unit[27]{M_\odot}$ signal model, as shown in the middle plots in Figures~\ref{fig:sign_fd} and~\ref{fig:sign_nd}.

No neutrino flavor changing oscillation effects were assumed when modeling the expected supernova neutrino flux.  This "no-oscillation" methodology is used in the template simulations to separate model and parameter dependant features from the baseline templates of the trigger.  To check for biases under this simplified assumption, we followed Ref.~\cite{Scholberg:2017czd} to take into account simple adiabatic flavor conversion from MSW resonances in the stellar medium.  This effect leads to a mixing, under the normal neutrino mass ordering hypothesis, or a swapping (under the inverted neutrino mass ordering, of $\bar\nu_e$ flux with $\bar\nu_x$ flux. The resulting $\bar\nu_e$ flux will have properties inherited from the $\bar\nu_x$ flux, resulting in fewer neutrinos but with higher neutrino energies than in the initial spectrum. 
This results in a two-sided effect on the expected number of events to be detected by NOvA.  It results in \unit[5]{\%} (\unit[17]{\%}) fewer neutrino interaction candidates for the brighter \unit[27]{$M_\odot$} progenitor model, and \unit[12]{\%} (\unit[40]{\%}) more neutrino candidates for the \unit[9.6]{$M_\odot$} progenitor in the case of normal (inverted) mass ordering. The enhancement of the signal for the smaller progenitor star is caused by the energy dependence of the interaction cross-section and the detection efficiency.  By using the ``no-oscillations'' methodology as a triggering template, NOvA has chosen a  conservative choice, which does not push the spectra to the two extremes as would be in the case of including oscillations.  This approach minimizes triggering biases with respect to the as yet unknown neutrino mass hierarchy and range of progenitor masses.

\begin{figure*}[!htb]
\centerline{
\includegr[scale=0.54]{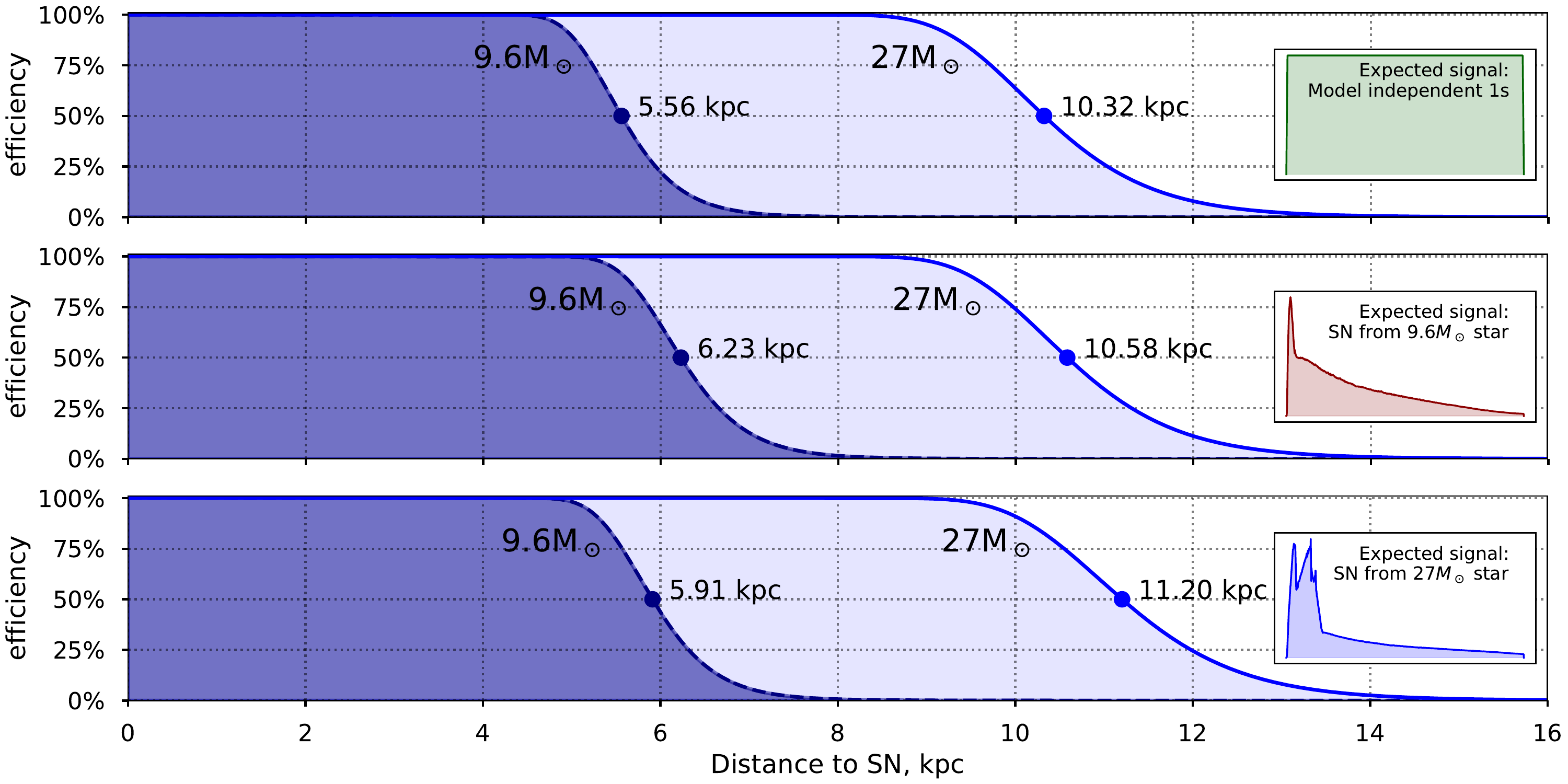}
}
\caption{NOvA's sensitivity to a galactic supernovae vs.\ distance by supernova model and expected signal shape for the \fardet.  The fraction of supernova occurrences for which the signal probability would be above threshold is plotted as a function of distance for the cases presented in Fig.~\ref{fig:sign_fd}.}
\label{fig:eff_distance_fd}
\end{figure*}

\begin{figure*}[!htb]
\centerline{
\includegr[scale=0.54]{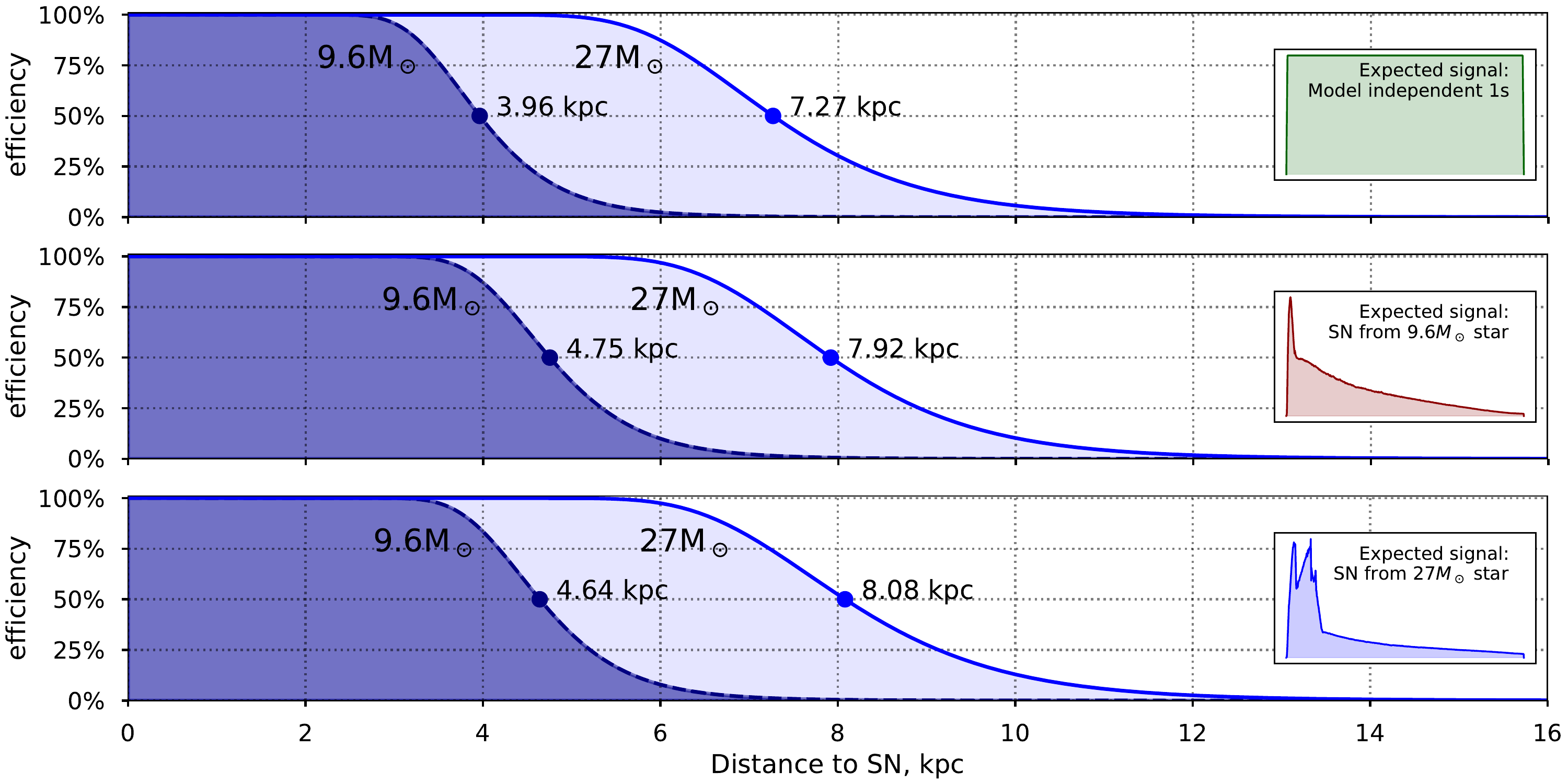}
}
\caption{NOvA's sensitivity to a galactic supernova vs.\ distance by supernova model and expected signal shape for the \neardet.   The fraction of supernova occurrences for which the signal probability would be above threshold is plotted as a function of distance for the cases presented in Fig.~\ref{fig:sign_nd}.}
\label{fig:eff_distance_nd}
\end{figure*}

\subsection{Probability of detecting the next galactic supernova}

\begin{figure*}[!htb]
\centerline{
\includegr[scale=0.54]{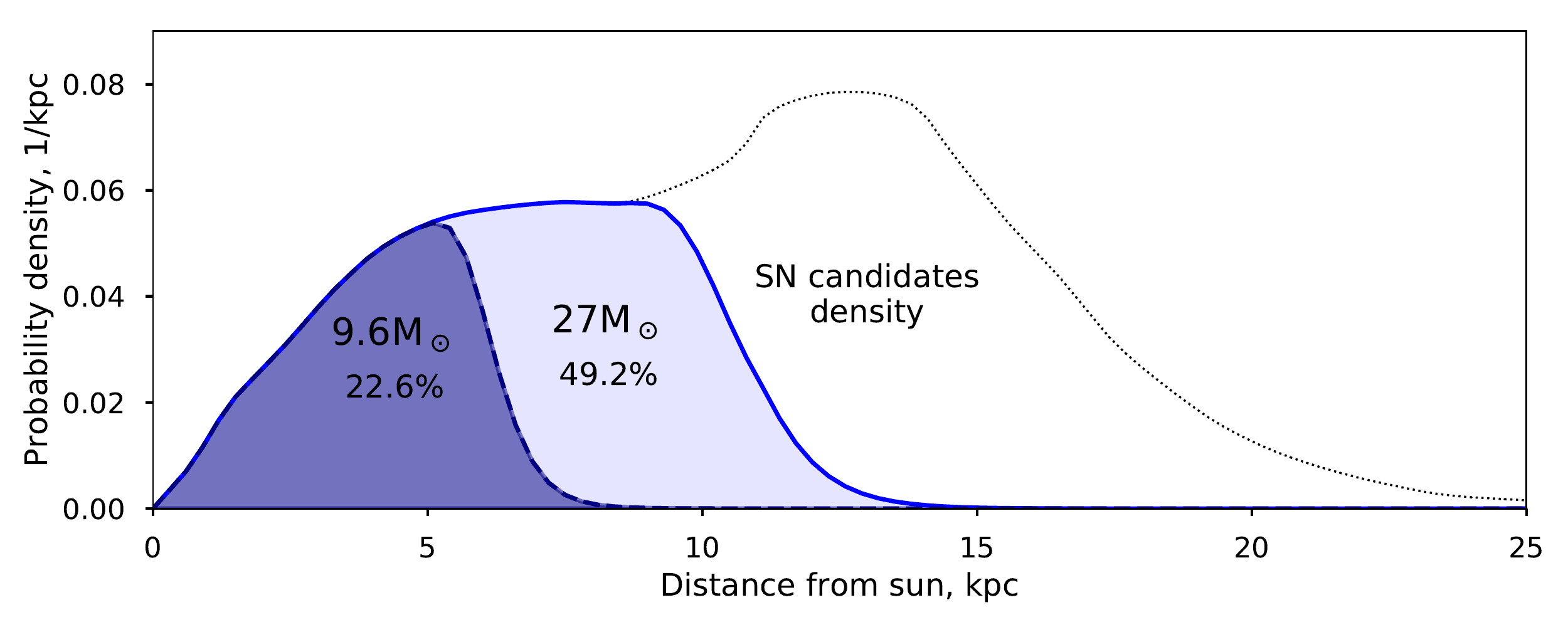}
}
\caption{The NOvA experiment's sensitivity to galactic supernova vs.\ distance.  The dark blue area is for a \unit[9.6]{M$_\odot$} progenitor, and the light blue area from a brighter \unit[27]{M$_\odot$} star: both models are from \cite{garching}.  The dotted curve for supernova progenitor density is from \cite{sn-density}. Percentage is the fraction of SN candidates covered by the trigger, for the two cases of  progenitor masses.}
\label{fig:sn_chance}
\end{figure*}

The probability of the NOvA detectors detecting a supernova for various distances (see Figs.~\ref{fig:eff_distance_fd} and  \ref{fig:eff_distance_nd}) can now be convoluted with the observed spatial distribution of potential supernova progenitors from Ref.~\cite{sn-density}.  This integral 
allows the calculation of the probability to detect the next galactic supernova with the NOvA supernova trigger system.
We have a 22.6\% and 49.2\% chance to detect a supernova from \unit[9.6]{M$_\odot$} and \unit[27]{M$_\odot$} progenitor stars, respectively, assuming the same spatial distribution for both progenitor masses. (see Fig.~\ref{fig:sn_chance}).

\section{Trigger system commissioning}
\label{sec:commissioning} 
\begin{figure*} \centering \includegr{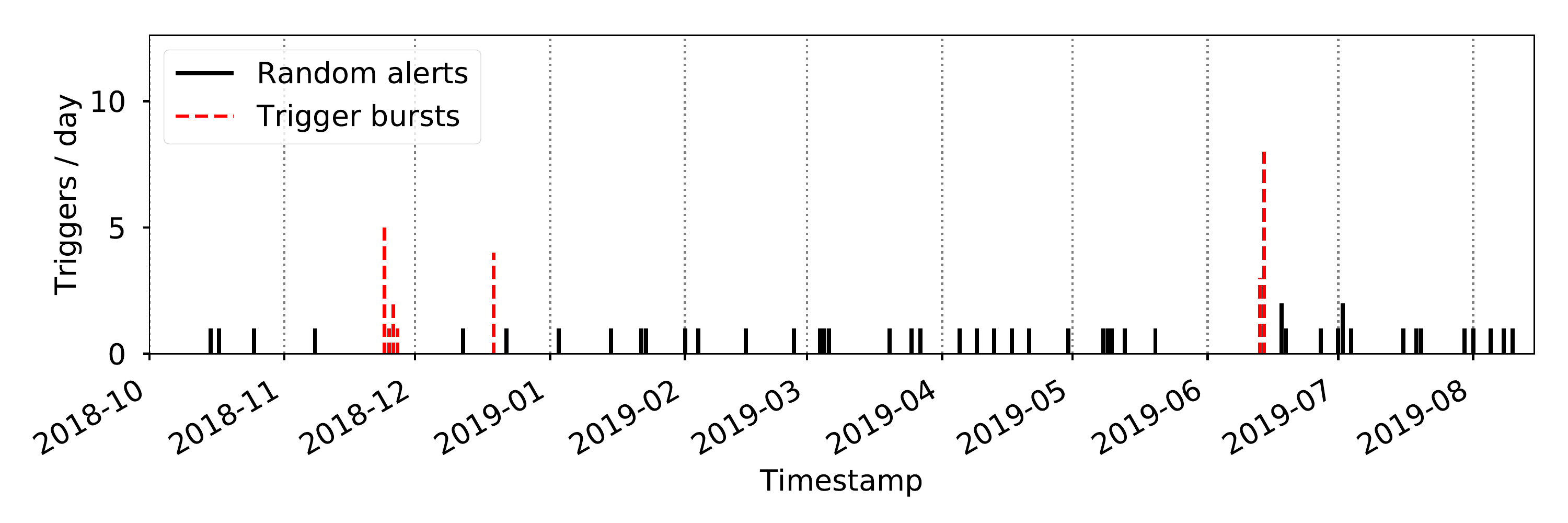}
    \caption{Time distribution of NOvA SN triggers issued by the
system on the \fardet during the commissioning period. Red dashed
lines show trigger bursts associated with unstable detector or readout
conditions as discussed in Sec.~\ref{sec:commissioning}.}
    \label{fig:triggering_timeline}
\end{figure*}

Since deploying the supernova triggering
system on November~1, 2017, several fixes and upgrades have been made to
reduce the rate of false triggering caused by instabilities in the
detector and readout conditions. During the 318 day commissioning
period from October~1, 2018 to~August 15, 2019, the NOvA \fardet
triggering system issued 71 supernova triggers. Each trigger requests
the data readout within \unit[45]{s}. In the case when another supernova
trigger is received within the same \unit[45]{s} window, the triggers
are merged into one request for a longer readout. The time
distribution of the issued triggers is shown in
Fig.~\ref{fig:triggering_timeline}.

Out of the observed 71 supernova triggers, 24 were concentrated in three trigger
bursts. These bursts were caused by readout instabilities,
described below.

\subsection{Partial detector data} 

It can take up to ten minutes after
the detector run restarts for the readout electronics to fully
synchronize, resulting in an underestimation of the background level
by the supernova triggering system. After the readout is synchronized,
the system is triggered by the perceived increase in detector
activity.  This false triggering mode is prevented by adding a filter
to remove incomplete data from the trigger pipeline.
    
\subsection{Noise channel map updates failure} 

As discussed in
Sec.~\ref{subsubsec:noisemap}, a map of noisy electronic channels is
updated every hour in order to adapt to changes in electronics
conditions. When such an update is not performed (in case the
computing process has stopped), a new noisy channel can emit a
periodic burst of noise hits, causing a false trigger every
ten minutes This problem was solved by setting up additional
monitoring of the noise map updating process.

\subsection{Triggering rate} 
\begin{figure*} \centering \includegr{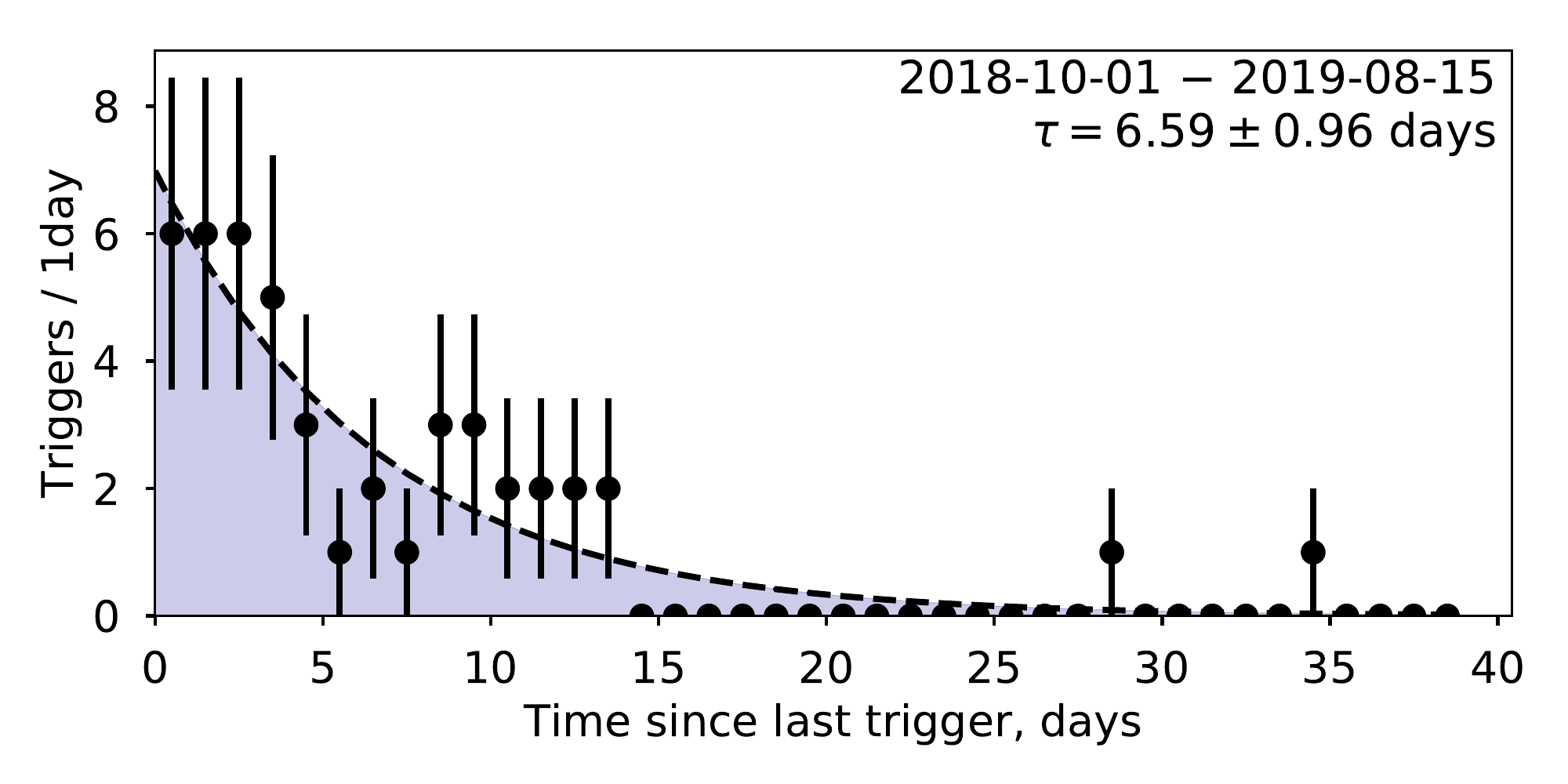}
    \caption{Distribution of the time intervals between subsequent SN
triggers at the NOvA Far Detector (excluding those due to "bursts"). The dashed line shows the
exponential fit of the distribution, with the time constant $\tau =
\unit[6.59\pm0.96$]{days}.}
    \label{fig:triggering_distr}
\end{figure*}

The remaining 47 supernova triggers
during the commissioning period are considered to be the result of the
expected random background fluctuations. Their average rate of
$1/(\unit[6.77 \pm 0.98]{days})$ is in agreement with the expected
one trigger per week. The distribution of time intervals between
subsequent triggers, as shown in Fig. \ref{fig:triggering_distr}, is
consistent with an exponential distribution with best fit average
rate of $1/(\unit[6.59\pm0.96]{days})$ (statistical errors).

\section{Summary}
The NOvA experiment's detectors are sensitive to the burst of neutrinos emitted by a core-collapse supernova in our galaxy through the detection of low energy ($\mathcal{O}$(\unit[10]{MeV})) positrons from inverse beta decay interactions.  The time structure and rate excess of these interactions allow for the ensemble of interactions to be distinguished from other cosmogenic backgrounds in the NOvA detectors.  The NOvA Near and Far detectors have different signal to noise ratios for these signatures due to the size of their overburdens and effective detector masses.  For both near and far detectors, data from the detectors are buffered by the DAQ systems and the data is reconstructed in near real time through low latency pattern recognition and background subtraction algorithms.  The rate of supernova neutrino-like events is assembled into a time series.  The time series is compared to template distributions from simulated supernovae to determine a likelihood metric which is used to compute the significance of the time series having originated from a supernova.  When the significance exceeds a threshold set within the trigger system, a trigger signal is generated which causes the DAQ system to initiate the recording of a \unit[45]{s} time window around the region of interest.  This region captures the supernova neutrino burst and activity both prior and subsequent to it, where the intervals are intended to yield a pre-sideband and a trailing tail that can extend out to black hole formation.

The significance threshold for this trigger process has been set to balance the trigger efficiency for detection of a supernova event with the false positive rate resulting from Poisson variation in the detector's activity.  Under these conditions, the NOvA supernova trigger is estimated from Monte Carlo modeling of supernova neutrino spectra to achieve a 50\% or higher trigger detection efficiency for a \unit[9.6]{M$_\odot$} progenitor's collapse out to a distance of \unit[6.2]{kpc}.  Similarly, the detectors achieve a 50\% or higher trigger detection efficiency for a brighter \unit[27]{M$_\odot$} progenitor out to a distance of \unit[10.6]{kpc}.  The resulting supernova system triggered 71 times between Oct. 1, 2018 and Aug. 15, 2019 with an average rate of  $1/(\unit[6.59\pm0.96]{days})$.  
During this period, there has been neither an optical observation nor a neutrino based observation from other experiments of a Milky Way supernova that is coincident with these triggered events.
We have found the triggering rate to be consistent with expected statistical fluctuations in the detector backgrounds and with periods of known instability in the detector readouts.

At distances greater than \unit[10]{kpc}, NOvA relies primarily on capturing a supernova neutrino burst through the use of an external trigger signal from the SNEWS coincidence network.  Future work on combining the time series obtained from both the near and far detectors into a joint significance will allow for greater sensitivity and reach for the self-triggering system.


\acknowledgments

 This document was prepared by the NOvA collaboration using the resources of the Fermi National Accelerator Laboratory (Fermilab), a U.S. Department of Energy, Office of Science, HEP User Facility. Fermilab is managed by Fermi Research Alliance, LLC (FRA), acting under Contract No. DE-AC02-07CH11359. This work was supported by the U.S. Department of Energy; the U.S. National Science Foundation; the Department of Science and Technology, India; the European Research Council; the MSMT CR, GA UK, Czech Republic; the RAS, RFBR, RMES, RSF, and BASIS Foundation, Russia; CNPq and FAPEG, Brazil; STFC, and the Royal Society, United Kingdom; and the state and University of Minnesota. 
 We are grateful for the contributions of the staffs of the University of Minnesota at the Ash River Laboratory and of Fermilab.
\clearpage

\bibliographystyle{JHEP}
\bibliography{snpaper}

\providecommand{\href}[2]{#2}\begingroup\raggedright\begin{thebibliography}{10}

\bibitem{Bionta:1987qt}
R.~M. Bionta et~al., \emph{{Observation of a neutrino burst in coincidence with
  supernova {SN1987A} in the {Large Magellanic Cloud}}}, {\emph{Phys. Rev.
  Lett.} {\bfseries 58} (1987) 1494}.

\bibitem{Hirata:1987hu}
{\scshape KAMIOKANDE-II} collaboration, \emph{{Observation of a neutrino burst
  from the supernova {SN1987A}}}, {\emph{Phys. Rev. Lett.} {\bfseries 58}
  (1987) 1490}.

\bibitem{Alekseev:1987ej}
E.~N. Alekseev, L.~N. Alekseeva, V.~I. Volchenko and I.~V. Krivosheina,
  \emph{{Possible detection of a neutrino signal on 23 {F}ebruary 1987 at the
  {B}aksan underground scintillation telescope of the {I}nstitute of {N}uclear
  {R}esearch}}, {\emph{JETP Lett.} {\bfseries 45} (1987) 589}.

\bibitem{Aglietta:1987it}
M.~Aglietta et~al., \emph{On the event observed in the {M}ont {B}lanc
  underground neutrino observatory during the occurrence of {S}upernova 1987a},
  {\emph{Europhys. Lett.} {\bfseries 3} (1987) 1315}.

\bibitem{snews}
P.~Antonioli et~al., \emph{{SNEWS: The Supernova Early Warning System}},
  \href{https://doi.org/10.1088/1367-2630/6/1/114}{\emph{New J. Phys.}
  {\bfseries 6} (2004) 114}
  [\href{https://arxiv.org/abs/astro-ph/0406214}{{\ttfamily
  astro-ph/0406214}}].

\bibitem{nova-tdr}
{\scshape NOvA} collaboration, \emph{{The {NOvA} Technical Design Report}}, .

\bibitem{numi_nim}
P.~Adamson et~al., \emph{{The {NuMI} Neutrino Beam}},
  \href{https://doi.org/10.1016/j.nima.2015.08.063}{\emph{Nucl. Instrum. Meth.}
  {\bfseries A806} (2016) 279}
  [\href{https://arxiv.org/abs/1507.06690}{{\ttfamily 1507.06690}}].

\bibitem{extrusion_nim}
R.~L. Talaga, J.~J. Grudzinski, S.~Phan-Budd, A.~Pla-Dalmau, J.~E. Fagan,
  C.~Grozis et~al., \emph{{PVC Extrusion Development and Production for the
  {NOvA} Neutrino Experiment}},
  \href{https://doi.org/10.1016/j.nima.2017.03.004}{\emph{Nucl. Instrum. Meth.}
  {\bfseries A861} (2017) 77}
  [\href{https://arxiv.org/abs/1601.00908}{{\ttfamily 1601.00908}}].

\bibitem{scint_nim}
S.~Mufson et~al., \emph{{Liquid Scintillator Production for the {NOvA}
  Experiment}}, \href{https://doi.org/10.1016/j.nima.2015.07.026}{\emph{Nucl.
  Instrum. Meth.} {\bfseries A799} (2015) 1}
  [\href{https://arxiv.org/abs/1504.04035}{{\ttfamily 1504.04035}}].

\bibitem{Norman:CHEP2015-6}
A.~Norman et~al., \emph{{Performance of the NOvA Data Acquisition and Trigger
  Systems for the full 14\,kT Far Detector}},
  \href{https://doi.org/10.1088/1742-6596/664/8/082041}{\emph{J. Phys. Conf.
  Ser.} {\bfseries 664} (2015) 082041}.

\bibitem{sn-review}
K.~Scholberg, \emph{{Supernova Neutrino Detection}},
  \href{https://doi.org/10.1146/annurev-nucl-102711-095006}{\emph{Ann. Rev.
  Nucl. Part. Sci.} {\bfseries 62} (2012) 81}
  [\href{https://arxiv.org/abs/1205.6003}{{\ttfamily 1205.6003}}].

\bibitem{livermore}
T.~Totani, K.~Sato, H.~E. Dalhed and J.~R. Wilson, \emph{Future detection of
  supernova neutrino burst and explosion mechanism},
  \href{https://doi.org/10.1086/305364}{\emph{The Astrophysical Journal}
  {\bfseries 496} (1998) 216}.

\bibitem{garching}
A.~Mirizzi, I.~Tamborra, H.-T. Janka, N.~Saviano, K.~Scholberg, R.~Bollig
  et~al., \emph{{Supernova Neutrinos: Production, Oscillations and Detection}},
  \href{https://doi.org/10.1393/ncr/i2016-10120-8}{\emph{Riv. Nuovo Cim.}
  {\bfseries 39} (2016) 1} [\href{https://arxiv.org/abs/1508.00785}{{\ttfamily
  1508.00785}}].

\bibitem{sndb}
K.~Nakazato, K.~Sumiyoshi, H.~Suzuki, T.~Totani, H.~Umeda and S.~Yamada,
  \emph{Supernova neutrino light curves and spectra for various progenitor
  stars: From core colllapse to proto-neutron star cooling},
  \href{https://doi.org/10.1088/0067-0049/205/1/2}{\emph{The Astrophysical
  Journal Supplement Series} {\bfseries 205} (2013) 2}.

\bibitem{StrumiaVissani}
A.~Strumia and F.~Vissani, \emph{Precise quasi-elastic neutrino/nucleon
  cross-section},
  \href{https://doi.org/https://doi.org/10.1016/S0370-2693(03)00616-6}{\emph{Physics
  Letters B} {\bfseries 564} (2003) 42 }.

\bibitem{NuElastic}
W.~J. Marciano and Z.~Parsa, \emph{Neutrino{\textendash}electron scattering
  theory}, \href{https://doi.org/10.1088/0954-3899/29/11/013}{\emph{Journal of
  Physics G: Nuclear and Particle Physics} {\bfseries 29} (2003) 2629}.

\bibitem{NConC12}
B.~Armbruster, I.~Blair, B.~Bodmann, N.~Booth, G.~Drexlin, V.~Eberhard et~al.,
  \emph{Measurement of the weak neutral current excitation
  $^{12}{C}(\nu_\mu,\nu_\mu^\prime) ^{12}{C}^{*}(1^+,1;15.1\,{MeV})$ at
  ${E}_{\nu_\mu}=29.8\,{MeV}$},
  \href{https://doi.org/https://doi.org/10.1016/S0370-2693(98)00087-2}{\emph{Physics
  Letters B} {\bfseries 423} (1998) 15 }.

\bibitem{genie}
C.~Andreopoulos, A.~Bell, D.~Bhattacharya, F.~Cavanna, J.~Dobson, S.~Dytman
  et~al., \emph{The {GENIE} neutrino {M}onte {C}arlo generator},
  \href{https://doi.org/https://doi.org/10.1016/j.nima.2009.12.009}{\emph{Nuclear
  Instruments and Methods in Physics Research Section A: Accelerators,
  Spectrometers, Detectors and Associated Equipment} {\bfseries 614} (2010) 87
  }.

\bibitem{Geant}
S.~Agostinelli et~al., \emph{{G}eant4 — a simulation toolkit},
  \href{https://doi.org/https://doi.org/10.1016/S0168-9002(03)01368-8}{\emph{Nuclear
  Instruments and Methods in Physics Research Section A: Accelerators,
  Spectrometers, Detectors and Associated Equipment} {\bfseries 506} (2003) 250
  }.

\bibitem{nova-detsim}
A.~Aurisano, C.~Backhouse, R.~Hatcher, N.~Mayer, J.~Musser, R.~Patterson
  et~al., \emph{The {NOvA} simulation chain}, {\emph{Journal of Physics:
  Conference Series} {\bfseries 664} (2015) 072002}.

\bibitem{nue_paper}
{\scshape NOvA} collaboration, \emph{{New Constraints on Oscillation Parameters
  from $\nu_e$ Appearance and $\nu_\mu$ Disappearance in the {NOvA}
  Experiment}}, \href{https://doi.org/10.1103/PhysRevD.98.032012}{\emph{Phys.
  Rev.} {\bfseries D98} (2018) 032012}
  [\href{https://arxiv.org/abs/1806.00096}{{\ttfamily 1806.00096}}].

\bibitem{nova_numu}
{\scshape NOvA} collaboration, \emph{{First measurement of muon-neutrino
  disappearance in {NOvA}}},
  \href{https://doi.org/10.1103/PhysRevD.93.051104}{\emph{Phys. Rev.}
  {\bfseries D93} (2016) 051104}
  [\href{https://arxiv.org/abs/1601.05037}{{\ttfamily 1601.05037}}].

\bibitem{nova_reco}
M.~Baird, J.~Bian, M.~Messier, E.~Niner, D.~Rocco and K.~Sachdev, \emph{Event
  reconstruction techniques in {NOvA}},
  \href{https://doi.org/10.1088/1742-6596/664/7/072035}{\emph{Journal of
  Physics: Conference Series} {\bfseries 664} (2015) 072035}.

\bibitem{timingProceeding}
{\scshape NOvA} collaboration, \emph{{The {NOvA} Timing System: A System for
  Synchronizing a Long Baseline Neutrino Experiment}},
  \href{https://doi.org/10.1088/1742-6596/396/1/012034}{\emph{J. Phys. Conf.
  Ser.} {\bfseries 396} (2012) 012034}.

\bibitem{nova_snews}
A.~Habig and J.~Zirnstein, \emph{Integration of the {{S}uper {N}ova {E}arly
  {W}arning {S}ystem} with the {NOvA} trigger}, {\emph{Journal of Physics:
  Conference Series} {\bfseries 664} (2015) 082015}.

\bibitem{Scholberg:2017czd}
K.~Scholberg, \emph{{Supernova Signatures of Neutrino Mass Ordering}},
  \href{https://doi.org/10.1088/1361-6471/aa97be}{\emph{J. Phys. G} {\bfseries
  45} (2018) 014002} [\href{https://arxiv.org/abs/1707.06384}{{\ttfamily
  1707.06384}}].

\bibitem{sn-density}
A.~{Mirizzi}, G.~G. {Raffelt} and P.~{Serpico}, \emph{Earth matter effects in
  supernova neutrinos: Optimal detector locations},
  \href{https://doi.org/10.1088/1475-7516/2006/05/012}{\emph{JCAP} {\bfseries
  0605} (2006) 012} [\href{https://arxiv.org/abs/astro-ph/0604300}{{\ttfamily
  astro-ph/0604300}}].

\end{thebibliography}\endgroup
\end{document}